\documentclass[twocolumn,english]{IEEEtran}
\usepackage[T1]{fontenc}
\usepackage[utf8]{inputenc}
\usepackage{float}
\usepackage{url}
\usepackage{graphicx}
\usepackage{setspace}

\usepackage{graphicx}
\usepackage{amssymb}
\usepackage{epstopdf}

\usepackage[ruled,vlined, linesnumbered]{algorithm2e}
\usepackage{amsmath, amsthm, amssymb}
\usepackage{shadethm}
\numberwithin{equation}{section}
\usepackage{rotating}
\usepackage{subfig}
\usepackage{float} 
\usepackage{soul} 
\usepackage{multirow}
\usepackage{todonotes}
\usepackage{wrapfig}
\usepackage{rotating}
\usepackage{xcolor}
\usepackage{fancyhdr}
\usepackage{array}
\usepackage{numprint}
\usepackage{xspace}
\usepackage[switch]{lineno}
\usepackage{xr}	

\graphicspath{{graphics/}}

\newcommand{\algo}[1]{\textsf{#1}}
\newcommand{\term}[1]{\emph{#1}}

\newcommand{\set}[1]{ \{ #1 \} }

\newcommand{\ie}{i.\,e.\xspace}
\newcommand{\eg}{e.\,g.\xspace}
\newcommand{\etal}{et al.\xspace}

\newcommand{\margindef}[1]{\marginpar{\textsf{{\scriptsize #1}}} } 

\makeatletter
\@addtoreset{paragraph}{section}
\makeatother

\newcount\shortyear\newcount\shorthour\newcount\shortminute
\shorthour=\time\divide\shorthour by 60\shortyear=\shorthour
\multiply\shortyear by 60\shortminute=\time\advance\shortminute by
-\shortyear
\shortyear=\year\advance\shortyear by -1900

\def\zeit{\number\shorthour:\ifnum\shortminute<10 0\number\shortminute
\else\number\shortminute\fi}

\theoremstyle{slplain}

\newtheorem{definition}{Def.}


\usepackage{babel}
\begin{document}
\author{
\IEEEauthorblockN{Christian L. Staudt and Henning Meyerhenke}\\ 
\IEEEauthorblockA{Faculty of Informatics, Karlsruhe Institute of Technology (KIT), Germany \\
\texttt{\{christian.staudt, meyerhenke\}@kit.edu}}}

\title{Engineering Parallel Algorithms for \\ Community Detection in Massive Networks}

\maketitle

\begin{abstract}
The amount of graph-structured data has recently experienced an enormous
growth in many applications. To transform such data into useful information,
fast analytics algorithms and software tools are necessary.
One common graph analytics kernel is disjoint community detection (or graph
clustering). Despite extensive research on heuristic solvers for this
task, only few parallel codes exist, although parallelism will be
necessary to scale to the data volume of real-world applications.
We address the deficit in computing capability by a flexible and extensible
community detection framework with shared-memory parallelism. Within
this framework we design and implement efficient parallel community detection heuristics:
A parallel label propagation scheme; the first large-scale parallelization of the 
well-known Louvain method, as well as an 
extension of the method adding refinement; and an ensemble scheme combining the above.
In extensive
experiments driven by the algorithm engineering paradigm, we identify
the most successful parameters and combinations of these algorithms.
We also compare our implementations with state-of-the-art competitors.
The processing rate of our fastest algorithm often reaches 50M edges/second.
We recommend the parallel Louvain method and our variant with refinement as both qualitatively strong
and fast. Our methods are suitable for massive data sets with billions of edges.\footnote{A 
preliminary version of this paper appeared in \textit{Proceedings of the 42nd International
Conference on Parallel Processing (ICPP 2013)}~\cite{sm2013ehpcdh}.}

\smallskip{}
\end{abstract}

\section{Introduction}
The data volume produced by electronic devices is growing at an enormous rate. Important classes of such 
data can be modeled by \emph{complex networks}, which are increasingly used to represent phenomena 
as varied as the WWW, social relations, and brain topology. 
The resulting graph data sets can easily reach billions of edges for many relevant applications.
Analyzing data of this volume in near
real-time challenges the state of the art in terms of hardware, software, and algorithms.
A particular challenge is not only the amount of data, but also its structure.
Complex networks have topological features which pose computational
challenges different from traditional HPC applications: In a \emph{scale-free} network, the presence of a few
high-degree nodes (hubs) among many low degree nodes 
generates load balancing issues. In a \emph{small-world} network, the entire graph can be visited in only a few hops
from any source node, which negatively affects cache performance.
To enable network analysis methods to scale, we need algorithmic methods that harness parallelism and 
apply specifically to complex networks.

In this work, we deal with the task of \emph{community detection} (also known as \emph{graph clustering}) in large
networks, \ie the discovery of dense subgraphs.
Among manifold applications, community detection has been used to counteract search engine rank manipulation~\cite{Schaeffer:2007vn},
to discover scientific communities in publication databases~\cite{Staudt:2012fk}, 
to identify functional groups of proteins in cancer research~\cite{jonsson2006cluster},
and to organize content on social media sites~\cite{gargi2011large}.
So far, extensive research on community detection in networks has given rise to a variety of definitions of what constitutes
a good community and a variety of methods for finding such communities,
many of which are described in surveys by Schaeffer~\cite{Schaeffer:2007vn}
and Fortunato~\cite{Fortunato201075}. Among these definitions, the
lowest common denominator is that a community is an internally
dense node set with sparse connections to the rest of the graph.
While it can be argued that communities can overlap, we restrict ourselves
to finding disjoint communities, i.e. a partition 
of the node set which uniquely assigns a node to a community.
The quality measure \emph{modularity}~\cite{girvan2002community}
formalizes the notion of a good community detection solution by comparing its \emph{coverage}
(fraction of edges within communities) to an expected value based on
a random edge distribution model which preserves the degree distribution. 
Modularity is not without flaws (like the \emph{resolution limit}~\cite{Fortunato:2007zr},
which can be partially overcome by different techniques~\cite{berry2011tolerating,lambiotte2010multi,DBLP:journals/jcss/MeoFFP14}) nor alternatives~\cite{Yang:2012fk}, but has
emerged as a well-accepted measure of community quality. This makes
modularity our measure of choice. While
optimizing modularity is NP-hard~\cite{modularity.np}, efficient
heuristics have been introduced which explicitly increase modularity. 


For graphs with millions to billions of edges, only (near) linear-time
community detection algorithms are practical. Several fast
 methods have been developed in recent years. Yet, there is
a lack of research in adapting these methods to take advantage
of parallelism. A recent attempt at assessing the state of the
art in community detection was the \emph{10th DIMACS
Implementation Challenge} \emph{on Graph Partitioning and Graph Clustering~}\cite{BaderMSW13graph}.
\emph{DIMACS} challenges are scientific competitions in which the
participants solve problems from a specified test set, with the aim
of high solution quality and high speed.
Only two of the 15 submitted implementations for modularity optimization relied on parallelism
and only very few could handle graphs with billions of edges in reasonable time. 

Accordingly, our objective is the development and implementation of
 parallel community detection heuristics which are able to handle massive
graphs quickly while also producing a high-quality solution. In
the following, the competitors of the \emph{DIMACS} challenge
will be used for a comparative experimental study. 
In the design of such heuristics, we necessarily trade off solution quality against running time.
The \emph{DIMACS} challenge also showed that there is no consensus 
on what running times are acceptable and how desirable an increase in
the third decimal place of modularity is. 
We therefore need to clarify our design goals as follows:
In the comparison with other proposed methods, we want to 
place our algorithms on the Pareto frontier so that they are not dominated, 
i.e. surpassed in speed \emph{and} quality at the same time. 
Secondly, we target a usage scenario: Our algorithms should be suitable
as part of interactive data analysis workflows,
performed by a data analyst operating a multicore workstation.
Networks with billions of edges should be processed in minutes rather than hours,
and the solution quality should be competitive with the results of
well-established sequential methods.


We implement three standalone parallel algorithms:
Label propagation~\cite{Raghavan:2007fk} is a simple procedure where nodes adopt the
community assignment (label) which is most frequent among their neighbors
until stable communities emerge. We implement a parallel
version of the approach as the \textsf{PLP} algorithm.
The Louvain method~\cite{Blondel:2008uq} is a multilevel
technique in which nodes are repeatedly moved to the community of a
neighbor if modularity can be improved. We are the first to present a parallel implementation of the method
for large inputs, named \textsf{PLM}. We also extend the method by adding a refinement
phase on every level, which yields the \textsf{PLMR} algorithm.
In addition to these basic algorithms, we also
implement a two-phase approach that combines them.
It is inspired by \emph{ensemble learning},
in which the output of several weak classifiers is combined to form
a strong one. In our case, multiple \emph{base algorithms} run in
parallel as an ensemble. Their solutions are then combined to form
the \emph{core communities}, representing the consensus of all base algorithms.
The graph is coarsened according to the core communities, and then
assigned to a single \emph{final algorithm}. Within
this extensible framework, which we call the \emph{ensemble preprocessing}
method (\textsf{EPP}), we apply \textsf{PLP} as base algorithms and
\textsf{PLMR} as the final algorithm.


With our shared-memory parallel implementation of community detection by label propagation (\textsf{PLP}), we provide an extremely fast basic
algorithm that scales well with the number of processors (considering the heterogeneous 
structure of the input). The processing
rate of \textsf{PLP} reaches 50M edges per second for large
graphs, making it suitable for massive data sets.
With \textsf{PLM}, we present the first parallel implementation of the
Louvain community detection method for massive inputs,
and demonstrate that it is both fast and qualitatively strong.
We show that solution quality can be further improved by extending the method with a
refinement phase on every level of the hierarchy, yielding the \textsf{PLMR} algorithm.
The \textsf{EPP} ensemble
algorithm can yield  a good quality-speed tradeoff on some instances when an even lower time to solution
is desired. In comparative experiments, our implementations perform well in comparison to
other state-of-the-art algorithms (Sec.~\ref{sub:Comparison} and~\ref{sub:Pareto}): 
Three of our algorithms are on the Pareto frontier.

Our community detection software framework, written in C++, is flexible, extensible, and supports
rapid iteration between design, implementation and testing required
for algorithm engineering~\cite{DBLP:conf/dagstuhl/2010ae}. In this
work, we focus on specific configurations of algorithms,
but future combinations can be quickly evaluated.
We distribute our community detection code as a component of \textit{NetworKit}~\cite{staudt2014networkit}, 
our open-source network analysis package, which is under continuous 
development.

%

\section{Related Work}

This section gives a short overview over related efforts. For a comprehensive
overview of community detection in networks, we refer the interested reader to 
aforementioned surveys~\cite{Schaeffer:2007vn,Fortunato201075}.
Recent developments and results are also covered by the \emph{10th DIMACS
Implementation Challenge}~\cite{BaderMSW13graph}.

Among efficient heuristics for community detection we can distinguish between those based on community agglomeration and those based on local node moves. Agglomerative algorithms successively merge pairs of communities so that an improvement with respect to community quality is achieved. In contrast, local movers search for quality gains which can be achieved by moving a node to the community of a neighbor.

A globally greedy agglomerative method known as \algo{CNM}~\cite{Clauset:2004ys} runs
in $O(md\log n)$ for graphs with $n$ nodes and $m$ edges, where
$d$ is the depth of the dendrogram of mergers and typically $d\sim\log n$.
Among the few parallel implementations competing in the \emph{DIMACS} challenge,
Fagginger Auer and Bisseling~\cite{Fagginger:2013gc} submitted an
agglomerative algorithm with an implementation for both the GPU (using
\emph{NVIDIA CUDA}) and the CPU (using \emph{Intel TBB}). The algorithm
weights all edges with the difference in modularity resulting from
a contraction of the edge, then computes a heavy matching $M$ and
contracts according to $M$. This process continues recursively with
a hierarchy of successively smaller graphs. The matching procedure
can adapt to star-like structures in the graph to avoid insufficient
parallelism due to small matchings. In the challenge, the CPU implementation
competed as \textsf{\textsc{CLU\_TBB}} and proved exceptionally fast.
Independently, Riedy \etal~\cite{Riedy:2013pc} developed a similar
method, which follows the same principle but does not provide the
adaptation to star-like structures. An improved implementation, labeled \textsf{\textsc{CEL}} in the following, corresponds to the 
description in~\cite{DBLP:conf/ipps/RiedyB13}.

Community detection by label propagation belongs to the class of local move heuristics.
It has originally been described by Raghavan
\etal~\cite{Raghavan:2007fk}. Several variants of the algorithm
exist, one of them (under the name \emph{peer pressure clustering})
is due to Gilbert \etal~\cite{gilbert2007high}. The latter use the algorithm as
a prototype application within a parallel toolbox that uses numerical
algorithms for combinatorial problems. Unfortunately, Gilbert \etal report
running times only for a different algorithm, which solves
a very specific benchmark problem and is not applicable in our context.
A variant of label
propagation by Soman and Narang~\cite{Soman:2011fk} for multicore
and GPU architectures exists, which seeks to improve quality by re-weighting
the graph.

A locally greedy multilevel-algorithm known as the \emph{Louvain method}~\cite{Blondel:2008uq}
combines local node moves with a bottom-up multilevel approach. 
Bhowmick and Srinivasan~\cite{bhowmick2013template} presented a previous parallel version of the algorithm.
According to their experimental results, our implementation is about four orders of magnitude
faster.
Noack and Rotta~\cite{Rotta:2011:MLS:1963190.1970376} evaluate similar
sequential multilevel algorithms, which combine agglomeration with refinement.

Ovelgönne and Geyer-Schulz~\cite{OvelgoenneG13ensemble} apply the
\emph{ensemble learning} paradigm to community detection. They develop
what they call the \emph{Core Groups Graph Clusterer } scheme,
which we adapt as the \emph{Ensemble Preprocessing} (\textsf{EPP})
algorithm. They also introduce an iterated scheme in which
the core communities are again assigned to an ensemble, creating a hierarchy
of solutions/coarsened graphs until quality does not
improve any more. Within this framework,
they employ \emph{ Randomized Greedy} (RG), a variant
of the aforementioned \algo{CNM} algorithm.
It avoids a loss in solution quality that arises from highly unbalanced
community sizes. The resulting \textsf{CGGC} algorithm emerged as the winner of the Pareto part of the DIMACS challenge, 
which related quality to speed according to specific rules.
Recently Ovelgönne~\cite{DBLP:conf/asunam/Ovelgonne13} presented a distributed
implementation (based on the big data framework \emph{Hadoop}) of
an ensemble preprocessing scheme using label propagation as a base
algorithm. This implementation processes a 3.3 billion edge web graph
in a few hours on a 50 machine Hadoop cluster~\cite[p.~73]{DBLP:conf/asunam/Ovelgonne13}.
(Our \emph{OpenMP}-based implementation of the similar \textsf{EPP}
algorithm requires only 4 \emph{minutes} on a shared-memory
machine with 16 physical cores.)

From an algorithmic perspective, disjoint community detection 
is related to graph partitioning (GP). Although the problems are different in important
aspects (unbalanced vs balanced blocks, unknown vs known number of blocks, different objectives), 
algorithms such as the Louvain method or \algo{PLMR} bear conceptual resemblance
to multilevel graph partitioners. Exploiting parallelism has been studied extensively for GP.
Several established tools are discussed in recent surveys~\cite{bichot2011graph,DBLP:journals/corr/BulucMSSS13},
most of them for machines with distributed memory. Often employed techniques are parallel matchings for coarsening
and parallel variants of Fiduccia-Mattheyses for local improvement. These techniques are at best partially helpful
in our scenario since vanilla matching-based coarsening is ineffective on complex networks and distributed-memory 
parallelism is not necessary for us. Related to our work is a recent study on multithreaded GP 
by LaSalle and Karypis~\cite{DBLP:conf/ipps/LasalleK13}, who explore the design space of multithreaded GP
algorithms. Their results provide interesting insights, but are not completely transferable
to our scenario.
Very recently they presented \textit{Nerstrand}~\cite{lasalle2014multi}, a fast parallel community detection algorithm based on modularity maximization and the multilevel paradigm, using different aggregation schemes. 
Our work on \algo{PLP} in this paper has also inspired a very promising parallel multilevel algorithm for partitioning massive complex networks~\cite{MeyerhenkeSS15parallel}.

We observe that most efficient disjoint community detection heuristics make use of agglomeration or local node moves, 
possibly in combination with multilevel or ensemble techniques. 
Both basic approaches can be adapted for parallelism, but this is currently the exception rather than the norm in our scenario. 
In this work we compare our own algorithms with the best currently available, sequential and parallel alike.

\section{Algorithms\label{sec:Algorithms}}
In this section we formulate and describe our parallel variants of
existing sequential community detection algorithms, as well as ensemble techniques
which combine them. Implementation details are also discussed. We use the following notation:
A graph, the abstraction of a network data set, is denoted as $G=(V,E)$
with a node set $V$ of size $n$ and
an edge set $E$ of size $m$. In the following, edges $\{u,v\}$
are undirected and have weights $\omega:E\to\mathbb{{R}}^{+}$. The
weight of a set of nodes is denoted as $\omega(E'):={\displaystyle {\textstyle \sum}_{\{u,v\}\text{\ensuremath{\in}E'}}
 \omega(u,v)}.$ A community detection solution $\zeta=\{C_{1},\dots,C_{k}\}$ is a partition of the
node set $V$ into disjoint subsets called communities. Equivalently,
such a solution can be understood as a mapping
where $\zeta(v)$ returns the community containing node $v$. For our
implementation, the nodes have consecutive integer identifiers $id(v)$
 and edges are pairs of node identifiers. A
solution is represented as an array indexed by integer
node identifiers and containing integer community identifiers.

\subsection{Parallel Label Propagation (\textsf{PLP})}

\paragraph{Algorithm} 
Community detection by label propagation, as originally introduced by Raghavan
\etal~\cite{Raghavan:2007fk}, extracts communities from a labelling
$V\to\mathbb{N}$ of the node set. Initially, each node is assigned
a unique label, and then multiple iterations over the node set are
performed: In each iteration, every node adopts the most frequent
label in its neighborhood (breaking ties arbitrarily).
Densely connected groups of nodes thus agree on a common label, and
eventually a globally stable consensus is reached, which usually corresponds
to a good solution for the network. Label propagation therefore finds
communities in nearly linear time: Each iteration takes $O(m)$ time,
and the algorithm has been empirically shown to reach a stable solution
in only a few iterations, though not mathematically proven to do so.
The number of iterations seems to depend more on the graph structure
than the size. More theoretical analysis is done by Kothapalli \etal~\cite{KothapalliPS13analysis}.
The algorithm can be described as a locally greedy \textit{coverage} maximizer, i.e.
 it tries to maximize the fraction of edges which are placed within communities
 rather than across.
 With its purely local update rule, it tends to get stuck in local optima of \textit{coverage} which implicitly are good solutions with respect to modularity: A label is likely to propagate through and cover a dense community,
 but unlikely to spread beyond bottlenecks. 
The local update rule 
 and the absence of global variables make label propagation well-suited for a parallel
implementation.

Algorithm~\ref{alg:PLP} denotes \textsf{PLP}, our parallel variant
of label propagation. We adapt the algorithm in a straightforward
way to make it applicable to weighted graphs. Instead of the most
frequent label, the \emph{dominant label }in the neighborhood is chosen,
i.e. the label $l$ that maximizes $\sum_{u\in N(v):\zeta(u)=l}\omega(v,u)$.
We continue the iteration until the number of nodes which changed their
labels falls below a threshold $\theta$.

\begin{algorithm}[h]
\caption{ \algo{PLP}: Parallel Label Propagation
\label{alg:PLP}
}
\begin{small}
\DontPrintSemicolon

\SetKwFor{ParallelFor}{parallel for}{}{endfor}
\SetKwData{Up}{updated}

\KwIn{graph $G = (V,E)$}
\KwResult{communities  $\zeta : V \to \mathbb{N}$}

\ParallelFor{$v \in V$}{
	$\zeta(v) \gets id(v)$ 
\\ }

$\Up \gets n$ \\
$V_{\textsf{active}} \gets V$  \\

\While{$\Up > \theta$}{
	$\Up \gets 0$ \\
	\ParallelFor{$v \in \{u \in V_{\textsf{active}}: \deg(u) > 0\}$ }{
		$l^\star \gets \arg\max_{l} \left \{ \sum_{u \in N(v): \zeta(u) = l} \omega(v, u) \right \}$ \\
		\eIf{$\zeta(v) \neq l^\star$}{
			$\zeta(v) \gets l^\star$ \\
			$\Up \gets \Up + 1$ \\
			$ V_{\textsf{active}} \gets V_{\textsf{active}} \cup N(v)$ \\
		}{
			$V_{\textsf{active}} \gets V_{\textsf{active}} \setminus \set{v}$
		}
	}
}

\Return{$\zeta$}
\end{small}
\end{algorithm}

\paragraph{Implementation} 

We make a few modifications to the original algorithm. 
In the original description~\cite{Raghavan:2007fk}, nodes are traversed
in random order. Since the cost of explicitly randomizing the node
order in parallel is not insignificant, we make this optional and
rely on some randomization through parallelism otherwise. We also
observe that forgoing randomization has a negligible effect on quality.
We avoid unnecessary computation by distinguishing between active
and inactive nodes. It is unnecessary to recompute the label weights
for a node whose neighborhood labels have not changed in the previous iteration.
Nodes which already have the heaviest label become inactive (Algorithm 1, line 14), and are
only reactivated if a neighboring node is updated (line 12). We restrict iteration
to the set of active nodes. Iterations are repeated until the number
of nodes updated falls below a threshold value. The motivation for
setting threshold values other than zero is that on some graph instances,
the majority of iterations are spent on updating only a very small
fraction of high-degree nodes (see Fig.~\ref{fig:lp-update} in the supplementary material for an example).
Since preliminary experiments have shown that time can be saved and
quality is not significantly degraded by simply omitting
these iterations, we set an update threshold of $\theta=n\cdot10^{-5}$.
Note that we do not use the termination criterion specified in~\cite{OvelgoenneG13ensemble}
as it does not lead to convergence on some inputs. The original criterion is to stop when
all nodes have the label of the relative majority in their neighborhood~\cite{Raghavan:2007fk}.

Label propagation can be parallelized easily by dividing the range
of nodes among multiple threads which operate on a common label array.
This parallelization is not free of race conditions, since by the
time the neighborhood of a node $u$ is evaluated in iteration $i$
to set $\zeta_{i}(u)$, a neighbor $v$ might still have the previous iteration's label $\zeta_{i-1}(v)$
or already $\zeta_{i}(v)$. The outcome thus depends on
the order of threads. However, these race conditions are acceptable
and even beneficial in an ensemble setting since they introduce random
variations and increase base solution diversity. This also corresponds
to \emph{asynchronous updating}, which has been found to avoid oscillation
of labels on bipartite structures~\cite{Raghavan:2007fk}. 
When dealing with scale-free networks whose degree distribution follows
a power law, assigning node ranges of equal size to each thread will
lead to load imbalance as computational cost depends on the node degree.
Instead of statically dividing the iteration among the threads, guided
scheduling (with\texttt{\ \#pragma omp parallel for schedule(guided)}) assigns node ranges
of decreasing size from a queue to available threads. This way
it can help to overcome load balancing issues, since threads processing large neighborhoods will receive fewer vertices in later phases of the dynamical assignment process. This introduces some overhead, but
we observed that guided scheduling is generally superior to static
parallelization for \textsf{PLP} and similar methods.

\subsection{Parallel Louvain Method (\textsf{PLM})\label{sub:Parallel-Louvain-Method-algo}}

\paragraph*{Algorithm}
The \emph{Louvain method} for community detection was first presented
by Blondel \etal~\cite{Blondel:2008uq}. It can be classified as
a locally greedy, bottom-up multilevel algorithm and uses modularity
as the objective function. In each pass, nodes are repeatedly moved
to neighboring communities such that the locally maximal increase in modularity
is achieved, until the communities are stable. Algorithm~\ref{alg:move} denotes this move phase.
Then, the graph is coarsened
according to the solution (by contracting each community into a supernode) and the procedure continues recursively,
forming communities of communities. Finally, the communities in the coarsest
graph determine those in the input graph by direct prolongation.

Computation of the objective function modularity is a central part of the algorithm.
Let $\omega(u,C):=\sum_{\{u,v\}:v\in C}\omega(u,v)$ be the weight of
all edges from $u$ to nodes in community $C,$ and define the \emph{volume}
of a node and a community as $vol(u):={\textstyle \sum}_{\{u,v\}:v\in N(u)}\omega(u,v)+2\cdot\omega(u,u)$
and $vol(C):=\sum_{u\in C}vol(u)$, respectively.
The modularity of a solution is defined as 
\begin{equation}
mod(\zeta, G) := \sum_{C \in \zeta} \left ( \frac{\omega(C)}{\omega(E)} - \frac{vol(C)^2}{4 \omega(E)^2} \right )
\end{equation}

Note that the change in modularity resulting from a node move can
be calculated by scanning only the local neighborhood of a node, because the difference in
modularity when moving node $u\in C$ to community $D$ is:

\begin{align*}
&\Delta mod(u,\ C\to D)=\frac{\omega(u,D\setminus\{u\})-\omega(u,C\setminus\{u\})}{\omega(E)} \\ &+\frac{ ( vol(C\setminus\{u\}) - vol(D\setminus\{u\}) ) \cdot vol(u)}{2\cdot\omega(E)^{2}}
\end{align*}

We introduce a shared-memory parallelization of the Louvain method
(\textsf{PLM}, Algorithm~\ref{alg:PLM}) in which node moves are
evaluated and performed in parallel instead of sequentially. This
approach may work on stale data so that a monotonous modularity increase
is no longer guaranteed. Suppose that during the evaluation of a possible
move of node $u$ other threads might have performed moves that affect
the $\Delta mod$ scores of $u$. In some cases this can lead to a
move of $u$ that actually decreases modularity. Still, such undesirable
decisions can also be corrected in a following iteration, which is
why the solution quality is not necessarily worse. Working only on
independent sets of vertices in parallel does not provide a solution
since the sets would have to be very small, limiting parallelism and/or
leading to the undesirable effect of a very deep coarsening hierarchy.
Concerns about termination turned out to be theoretical for our set
of benchmark graphs, all of which can be successfully processed with
\textsf{PLM}.
The community size resolution produced by \textsf{PLM} can be varied through a parameter $\gamma$ in the range $[0, 2 m]$, $0$ yielding a single community, $1$ being standard modularity and $2m$ producing singletons. Tuning this parameter is a possible practical remedy~\cite{lambiotte2010multi} against modularity's resolution limit.

\begin{algorithm}[h]
\caption{\algo{move}: Local node moves for modularity gain
\label{alg:move}
}
\begin{small}
\DontPrintSemicolon

\SetKwFor{ParallelFor}{parallel for}{}{endfor}

\KwIn{graph $G = (V,E)$, communities $\zeta : V \to \mathbb{N}$}
\KwResult{communities  $\zeta : V \to \mathbb{N}$}

\Repeat{$\zeta$ stable}{

	\ParallelFor{$u \in V$} {
		$\delta \gets \max_{v \in N(u)} \left \{ \Delta mod(u, \zeta(u) \rightarrow \zeta(v)) \right \} $\\
		$C \gets \zeta(\arg\max_{v \in N(u)} \left \{ \Delta mod(u, \zeta(u) \rightarrow \zeta(v)) \right \} ) $\\
		\If{$\delta > 0$}{
			$\zeta(u) \leftarrow C$ \\
		}
	}
}
\Return{$\zeta$}
\end{small}
\end{algorithm}

\begin{algorithm}[h]
\caption{\algo{PLM}: Parallel Louvain Method 
\label{alg:PLM}
}
\begin{small}
\DontPrintSemicolon

\SetKwFor{ParallelFor}{parallel for}{}{endfor}
\SetKwData{Up}{up}

\KwIn{graph $G = (V,E)$}
\KwResult{communities  $\zeta : V \to \mathbb{N}$}

$\zeta \gets \zeta_{\textsf{singleton}}(G)$ \\

$\zeta \gets$ \textsf{move}($G$, $\zeta$) \\
\If{$\zeta$ changed}{
	$[G', \pi] \gets$ \textsf{coarsen}$(G, \zeta)$ \\
	$\zeta' \gets \textsf{PLM}(G')$ \\
	$\zeta \gets$ \textsf{prolong($\zeta'$, $G$, $G'$, $\pi$)} \\
}

\Return{$\zeta$}
\end{small}
\end{algorithm}

\paragraph*{Implementation}
The main idea of \textsf{PLM} (Algorithm~\ref{alg:PLM}) is to parallelize both the node move
phase and the coarsening phase of the Louvain method.
 Since the computation of the
$\Delta mod$ scores is the most frequent operation, it needs to be
very fast. We store and update some interim values, which is
not apparent from the high-level pseudocode in Algorithm~\ref{alg:PLM}.
An earlier implementation associated with each node a map in which the 
edge weight to neighboring communities was stored and updated when node moves
occurred. A lock for each vertex $v$ protected all read and write
accesses to $v$'s map since \texttt{std::map} is not thread-safe.
Meant to avoid redundant computation, we later discovered that this introduces too
much overhead (map operations, locks).
Recomputing the weight to neighbor communities each time 
a node is evaluated turned out to be faster.
 The current implementation only stores and updates the volume of each community.
An additional optimization to the \textsf{PLM} implementation eliminated the overhead associated with using an \texttt{std::map} to store for each node the weights of edges leading to neighboring communities.
the mechanism was replaced by one \texttt{std::vector} for each of the $p$ threads, leading to an acceleration of a factor of 2 on average, at the cost of a memory overhead of $O(p\cdot n)$.
The former version (referred to as \textsf{PLM*}) can still be used optionally under tighter memory constraints.

Graph coarsening according to communities is performed in a straightforward
way such that the nodes of a community in $G$ are aggregated to a single
node in $G'$. An edge between two nodes in $G'$ receives as weight
the sum of weights of inter-community edges in $G$, while self-loops
preserve the weight of intra-community edges. A mapping $\pi$ of nodes in the fine graph to nodes in the coarse graph is also returned. 
In earlier versions of \textsf{PLM}, the graph coarsening phase proved to be a major sequential bottleneck. 
We address this problem with a parallel coarsening scheme: 
Each thread first scans a portion of the edges in $G$ and constructs a coarse graph $G'_t$ of its own. These partial graphs are then combined into $G'$ by processing each node of $G'$ in parallel and merging the adjacencies stored in each $G'_t$.

\subsection{Parallel Louvain Method with Refinement (\textsf{PLMR})}

Following up on the work by Noack and Rotta on multilevel techniques and refinement heuristics~\cite{Rotta:2011:MLS:1963190.1970376}, we extend the Louvain method by an additional \textsf{move} phase after each prolongation. This makes it possible to re-evaluate node assignments in view of the changes that happened on the next coarser level, giving additional opportunities for modularity improvement at the cost of additional iterations over the node set in each level of the hierarchy. We denote the method and implementation as \textsf{PLMR} for \textit{Parallel Louvain Method with Refinement}. We present a recursive implementation in Algorithm~\ref{alg:PLMR} which uses the same concepts as \textsf{PLM}.

\begin{algorithm}[h]
\caption{\algo{PLMR}: Parallel Louvain Method with Refinement
\label{alg:PLMR}
}
\begin{small}
\DontPrintSemicolon

\SetKwFor{ParallelFor}{parallel for}{}{endfor}
\SetKwData{Up}{up}

\KwIn{graph $G = (V,E)$}
\KwResult{communities  $\zeta : V \to \mathbb{N}$}

$\zeta \gets \zeta_{\textsf{singleton}}(G)$ \\
$\zeta \gets \textsf{move}(\zeta, G)$ \\

\If{$\zeta$ changed}{
	$[G', \pi] \gets \textsf{coarsen}(G, \zeta)$ \\
	$\zeta' \gets \textsf{PLMR}(G')$ \\
	$\zeta \gets$ \textsf{prolong($\zeta'$, $G$, $G'$, $\pi$)} \\
	$\zeta \gets \textsf{move}(\zeta, G)$
}
\Return{$\zeta$}
\end{small}
\end{algorithm}

\subsection{Ensemble Preprocessing (\textsf{EPP})}

In machine learning, \emph{ensemble learning} is a strategy
in which multiple \emph{base classifiers} or \emph{weak classifiers}
are combined to form a strong classifier. Classification in this context can be
understood as deciding whether a pair of nodes should belong to the
same community. We follow this general idea, which has been applied
successfully to graph clustering before~\cite{OvelgoenneG13ensemble}.
Subsequently, we describe an ensemble techniques \textsf{EPP}.
We also briefly describe algorithms for combining multiple base solutions.

\begin{algorithm}[h]
\caption{\algo{EPP}: \term{Ensemble Preprocessing}}
\label{alg:EPP}

\begin{small}

\SetKwFor{ParallelFor}{parallel for}{}{endfor}

\KwIn{graph $G = (V,E)$, ensemble size $b$}
\KwResult{communities  $\zeta: V \to \mathbb{N}$}

\ParallelFor{$i \in [1,b]$}{
	$\zeta_i \gets \algo{Base}_i(G)$ \\
}

$\bar{\zeta} \gets \algo{combine}(\zeta_1, \dots, \zeta_b)$\\
$G', \pi \gets \algo{coarsen}(G, \bar{\zeta})$ \\

$\zeta' \gets \algo{Final}(G')$ \\
$\zeta \gets \algo{prolong}(\zeta', G, G', \pi) $\\

\Return{$\zeta$}

\end{small}
\end{algorithm}

In a preprocessing step,
assign $G$ to an ensemble of base algorithms. The graph is then coarsened
according to the \emph{core communities} $\bar{{\zeta}}$, which represent
the consensus of the base algorithms. Coarsening reduces the problem size considerably,
and implicitly identifies the contested and the unambiguous parts of
the graph. After the preprocessing phase, the coarsened graph $G'$
is assigned to the final algorithm, whose result is applied to the
input graph by prolongation. 
Our implementation of the ensemble technique \textsf{EPP}
is agnostic to the base and final algorithms and can be instantiated
with a variety of such algorithms. 
We instantiate the scheme with \textsf{PLP}
as a base algorithm and \textsf{PLMR} as the final algorithm. Thus
we achieve massive nested parallelism with several parallel \textsf{PLP}
instances running concurrently in the first phase, and proceed in
the second phase with the more expensive but qualitatively superior
 \textsf{PLMR}. This constitutes the \textsf{EPP} algorithm (Algorithm~\ref{alg:EPP}).
We write \textsf{EPP($b$, Base, Final)} to indicate the size of the ensemble $b$ and the types of
base and final algorithm.

\paragraph*{Implementation}

A consensus of $b>1$ base algorithms is formed by combining the base
solutions $\zeta_{i}$ in the following way: Only if a pair of nodes
is classified as belonging to the same community in every $\zeta_{i},$
then it is assigned to the same community in the core communities $\bar{{\zeta}}.$
Formally, for all node pairs $u,v\in V$:

\begin{equation}
\forall i\in[1,b]\ \zeta_{i}(u)=\zeta_{i}(v)\quad\iff\quad\bar{{\zeta}}(u)=\bar{{\zeta}}(v).\label{eq:core-groups}
\end{equation}

We introduce a highly parallel combination algorithm based on \emph{hashing}. With
a suitable hash function $h(\zeta_{1}(v),\dots,\zeta_{b}(v))$, the
community identifiers of the base solutions are mapped to a new identifier
$\bar{{\zeta}}(v)$ in the core communities. Except for unlikely hash
collisions, a pair of nodes will be assigned to the same community only
if the criterion above is satisfied. We use a relatively simple function called \textsf{djb2}
due to Bernstein,%
\footnote{hash functions: \url{http://www.cse.yorku.ca/~oz/hash.html}%
} which appears sufficient for our purposes.
The use of a $b$-way
hash function is fast due to a high degree
of parallelism.

\section{Implementation and Experimental Setup\label{sec:ExperimentalSetup}}

\subsection{Framework and Settings}

The language of choice for all implementations is C++ according to the
C++11 standard, allowing
us to use object-oriented and functional programming concepts while
also compiling to native code. We implemented all algorithms on top
of a general-purpose adjacency array graph data structure. 
Basically, it represents the adjacencies of each node by storing them in an
\texttt{std::vector}, allowing for efficient insertions and deletions of nodes and edges.
A high-level
interface encapsulates the data structure and enables a clear and
concise notation of graph algorithms. In particular, our interface
conveniently supports parallel programming through parallel node and
edge iteration methods which receive a function
and apply it to all elements in parallel. Parallelism is achieved
in the form of loop parallelization with \emph{OpenMP}, using the
\texttt{parallel for} directive with \texttt{schedule(guided)} where
appropriate for improved load balancing.

We publish our source code under a permissive free software license to
encourage reproduction, reuse and contribution by the community.
Implementations of all community detection algorithms mentioned are part of \textit{NetworKit}~\cite{staudt2014networkit},
our growing toolkit for network analysis.\footnote{\textit{NetworKit}: \url{https://networkit.iti.kit.edu/}%
} The software combines fast parallel algorithms written in C++ with 
an interactive Python interface for flexible and interactive data analysis workflows.

For representative experiments we average quality and speed values
over multiple runs in order to compensate for fluctuations. Table~\ref{tab:Platform}
provides information on the multicore platform used for all experiments.

\begin{table}[h]
\begin{center}
\begin{small}

    \begin{tabular}{|l|l|}
    \hline
    ~ & phipute1.iti.kit.edu \\
    \hline
    compiler & gcc 4.8.1                                                          \\ \hline
    CPU      & 2 x 8 Cores: Intel(R) Xeon(R) \\ 
    & E5-2680 0 @ 2.70GHz, 32 threads \\ \hline
    RAM       & 256 GB \\  \hline
        OS       & SUSE 13.1-64                                                       \\ \hline
    \end{tabular}
    \end{small}

\end{center}
\caption{Platform for experiments}
\label{tab:Platform}
\end{table}

\vspace{-4ex}

\subsection{Networks}
\label{sub:Graphs}
We perform experiments on a variety of graphs from different categories
of real-world and synthetic data sets. Our focus is on real-world complex networks, but to add variety some
non-complex and synthetic instances are included as well.
The test set includes web graphs ({\small \texttt{uk-2002}, \texttt{eu-2005}, \texttt{in-2004}, \texttt{web-BerkStan}}),
internet topology networks ({\small \texttt{as-22july06}, \texttt{as-Skitter}, \texttt{caidaRouterLevel}}), social networks ({\small\texttt{soc-LiveJournal, fb-Texas84, com-youtube, wiki-Talk, soc-pokec, com-orkut}}), scientific coauthorship
networks ({\small\texttt{coAuthorsCiteseer}, \texttt{coPapersDBLP}}), a connectome graph (\texttt{con-fiber\_big}), a street network ({\small\texttt{europe-osm}}) and synthetic graphs ({\small\texttt{G\_n\_pin\_pout}, \texttt{kron\_g500-simple-logn20}, \texttt{hyperbolic-268M}}).
Therefore, we cover a range of graph-structural properties.
Real-world complex networks are heterogeneous data sets,
which makes it impossible to pick an ideal or generic instance from which to generalize.
Our main test set is chosen such that it can be handled by competing codes as well.
It contains 20 networks from different
domains. With this test set we aim for generalizable results.
Note that the achievable modularity for a network depends on its size and
inherent community structure, which may or may not be distinctive, and varies widely
among the instances.
The majority of test networks are taken from the collection compiled
for the \emph{10th DIMACS Implementation Challenge}\footnote{\emph{DIMACS} collection: \url{http://www.cc.gatech.edu/dimacs10/downloads.shtml}%
} as well as the  \emph{Stanford
Large Network Dataset Collection}\footnote{\emph{Stanford} collection: \url{http://snap.stanford.edu/data/index.html}%
}
and  are freely available on the web.
They are undirected, unweighted graphs. 
Table~\ref{tab:Graphs} gives an overview over graph sizes as well as some structural features: A high maximum node degree (maxdeg) indicates possible load balancing issues. The number of connected components (comp) points to isolated single nodes or small groups of nodes. A high average local clustering coefficient (lcc) is an indicator for the presence of dense subgraphs.
We evaluate solution quality and running time for
all of our own algorithms as well as several relevant competitors on this set.
For those algorithms that can process in reasonable time the largest real-world graph
available to us, a web graph of the .uk domain with $m\approx3.3\cdot10^{9}$,
we add further experiments (see Section~\ref{sub:one-more}).
To measure strong scaling, we run our parallel algorithms on
this web graph.

\begin{table}[tb]
\rotatebox{0}{
\hspace{-3ex}
\begin{scriptsize}
\begin{tabular}{|l|r|r|r|r|r|r|}
\hline
graph &         n &          m &  maxdeg &    comp &     lcc \\
\hline
as-22july06 &     22963 &      48436 &    2390 &       1 &  0.3493 \\ \hline
G\_n\_pin\_pout &    100000 &     501198 &      25 &       6 &  0.0040 \\ \hline
caidaRouterLevel &    192244 &     609066 &    1071 &     308 &  0.2016 \\ \hline
coAuthorsCiteseer &    227320 &     814134 &    1372 &       1 &  0.7629 \\ \hline
fb-Texas84 &     36371 &    1590655 &    6312 &       4 &  0.1985 \\ \hline
com-youtube &   1157828 &    2987624 &   28754 &   22939 &  0.1725 \\ \hline
wiki-Talk &   2394385 &    4659565 &  100029 &    2555 &  0.1991 \\ \hline
web-BerkStan &    685231 &    6649470 &   84230 &     677 &  0.6343 \\ \hline
as-Skitter &   1696415 &   11095298 &   35455 &     756 &  0.2930 \\ \hline
in-2004 &   1382908 &   13591473 &   21869 &     134 &  0.7013 \\ \hline
coPapersDBLP &    540486 &   15245729 &    3299 &       1 &  0.8111 \\ \hline
eu-2005 &    862664 &   16138468 &   68963 &       1 &  0.6509 \\ \hline
soc-pokec &   1632804 &   22301964 &   14854 &       2 &  0.1223 \\ \hline
soc-LiveJournal &   4847571 &   43369619 &   20334 &    1876 &  0.3667 \\ \hline
kron\_g500-simple... &   1048576 &   44619402 &  131503 &  253380 &  0.2096 \\ \hline
con-fiber\_big &    591428 &   46374120 &    5166 &     727 &  0.6024 \\ \hline
europe-osm &  50912018 &   54054660 &      13 &       1 &  0.0012 \\ \hline
com-orkut &   3072627 &  117185083 &   33313 &     187 &  0.1735 \\ \hline
uk-2002 &  18520486 &  261787258 &  194955 &   38359 &  0.6892 \\ \hline
hyperbolic-268M &   6710886 &  268851810 &   71585 &       1 &  0.7895 \\ \hline
uk-2007-05              & 105896555  & 3301876564 &   975419                &   756936             & 0.743    \\ \hline
\end{tabular}
\end{scriptsize}
}
\caption{Overview of graphs used in experiments}
\label{tab:Graphs}
\end{table}

\section{Experiments and Results}

In this section we report on a representative subset of our experimental
results for our different parallel algorithms, as well as competing codes.
Figures~\ref{fig:performance-ours} and~\ref{fig:performance-theirs} (as well as Figures~\ref{fig:performance-theirs2} and~\ref{fig:performance-ours2} in the supplementary material)
show running time and quality differences broken down by the networks of our test set. 
The bars of the charts are in ascending order of graph size.
We have selected a diverse test set and show results for each network.
The Pareto evaluation (Section~\ref{sub:Pareto}) then aims to condense this into a single performance score.

\subsection{Parallel Label Propagation (\textsf{PLP})}

\begin{figure}[h]
\begin{centering}
\includegraphics[height=3.25cm]{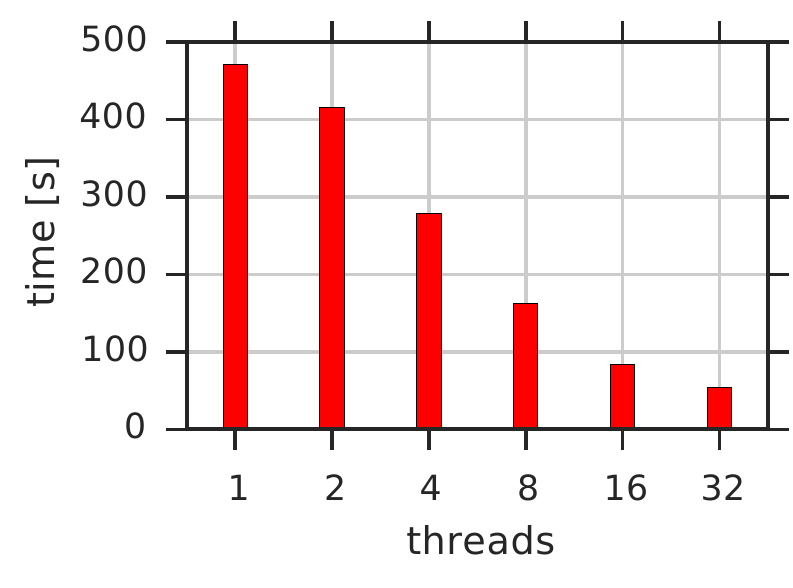}
\includegraphics[height=3.25cm]{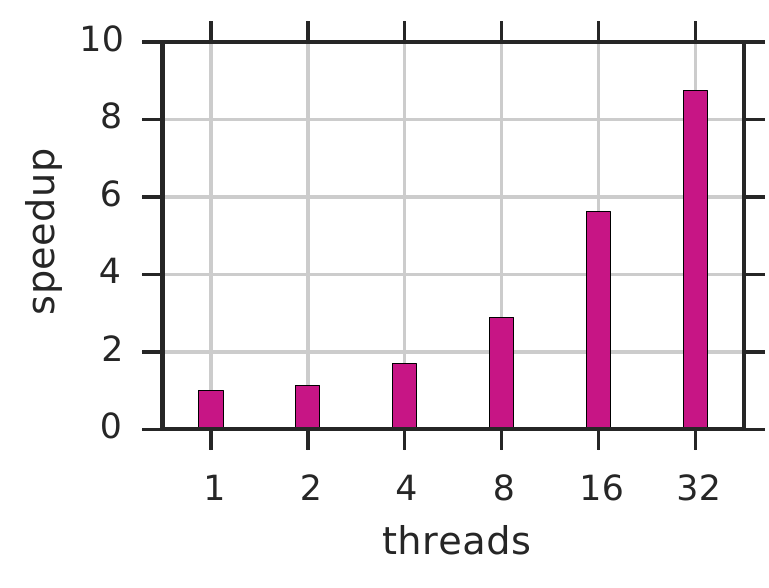}

\caption{\textsf{\label{fig:plp-strongscaling}PLP} strong scaling on the \texttt{uk-2007-05} web graph}
\end{centering}
\end{figure}

\textsf{PLP} is extremely fast and able to handle the large graphs
easily.
The ``weak classifier'' \textsf{PLP}
is nonetheless able to detect an inherent community structure and
produce a solution with reasonable modularity values, although it
cannot distinguish communities in a Kronecker graph, which has a very
weak community structure.
To demonstrate
strong scaling behavior, we apply \textsf{PLP} to the large \texttt{uk-2007-05}
web graph and increase the number of threads from 1 to 32 (Figure~\ref{fig:plp-strongscaling}).
(Weak scaling results
on \textsf{PLP} and \textsf{PLM} are shown in Figure~\ref{fig:plm-weakscaling}.) 
A speedup of about factor 8 is achieved when scaling from 1 to 32 threads.
Note that we have only 16 physical cores and the step from 16 to 32 threads implies 
hyperthreading, so that a lower speedup is expected.
Our results indicate that \textsf{PLP} can benefit from increased parallelism.
Figure~\ref{fig:plp-iter} in the supplementary material breaks running times down by iteration, showing
that the vast majority of time is spent in the first couple of iterations.

\subsection{Parallel Louvain Method (\textsf{PLM})\label{sub:PLM-Results}}

\begin{figure}[h]
\begin{center}
\includegraphics[height=3.2cm]{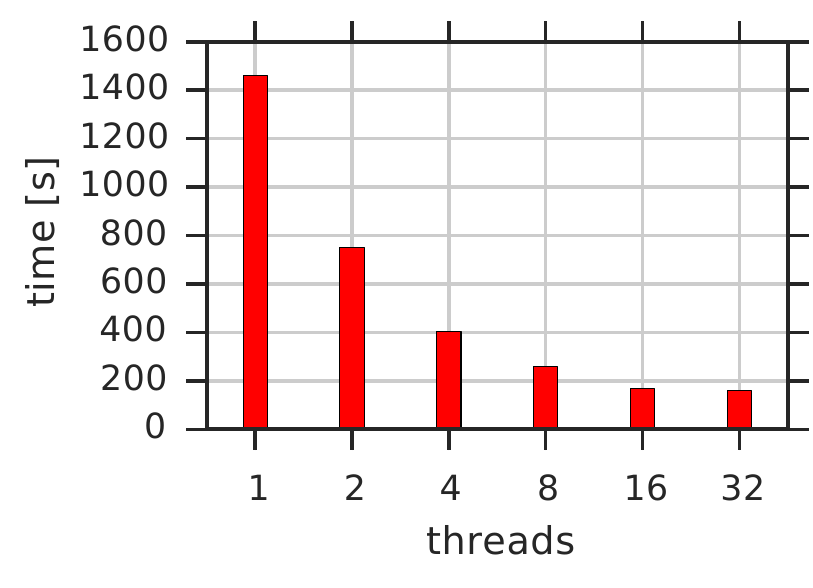}
\includegraphics[height=3.2cm]{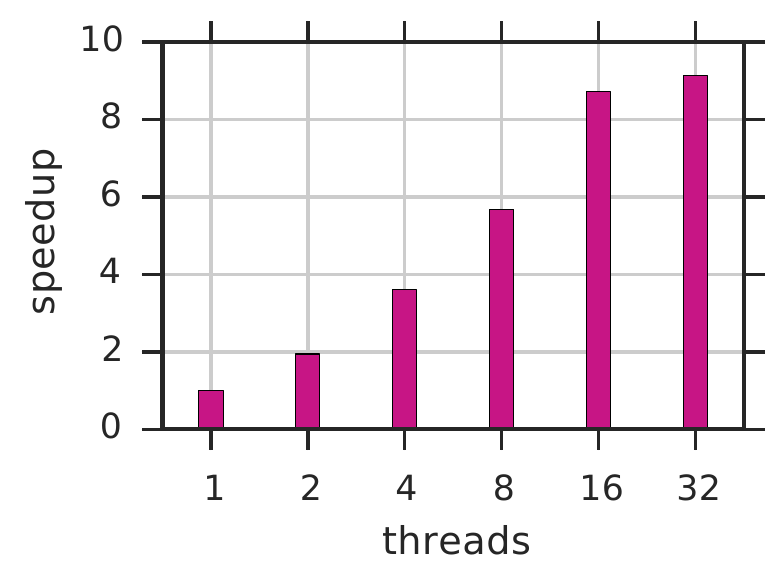}

\caption{\textsf{\label{fig:plm-scaling}PLM} strong scaling on the \texttt{uk-2007-05} web graph}
\end{center}
\end{figure}

\begin{figure}[!p]
  \begin{center}
      \subfloat[\textsf{PLM} : absolute quality and speed serve as baseline for comparison \label{fig:plm-baseline}]{%
\includegraphics[width=.45\columnwidth]{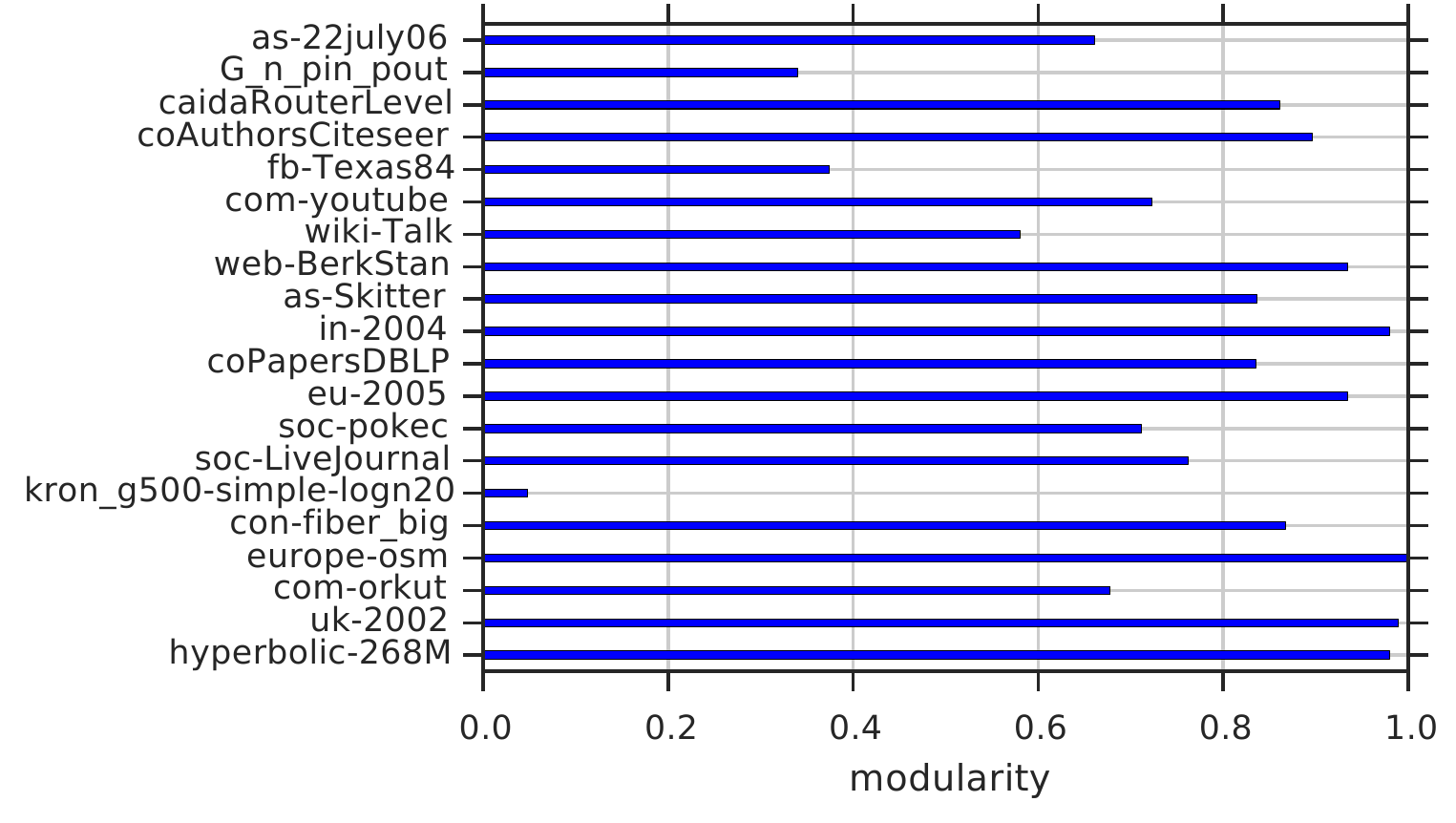}
\includegraphics[width=.45\columnwidth]{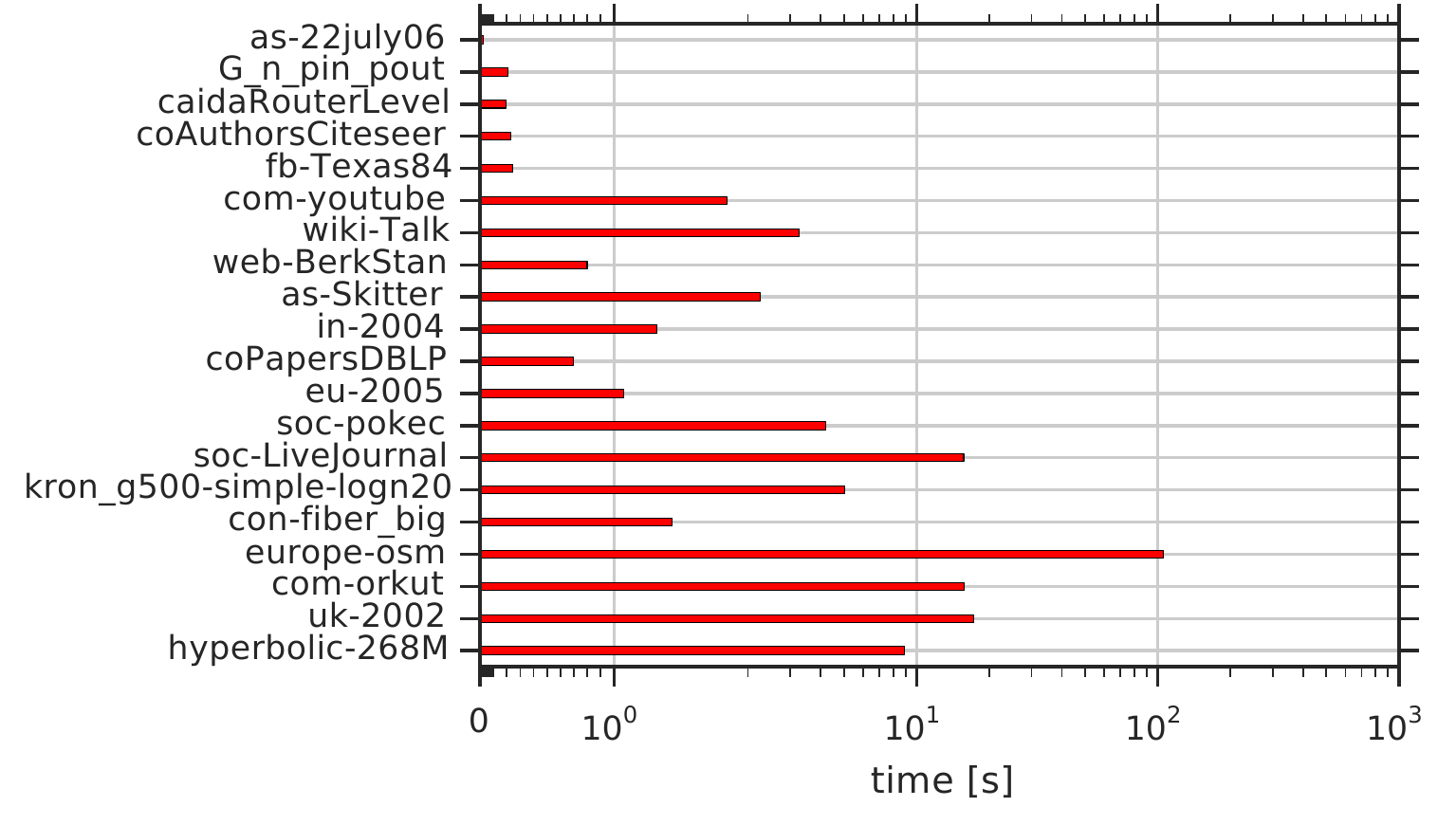}
    }
    \hfill
    \subfloat[\textsf{PLP} \label{fig:plp-rel}]{%
\includegraphics[width=.45\columnwidth]{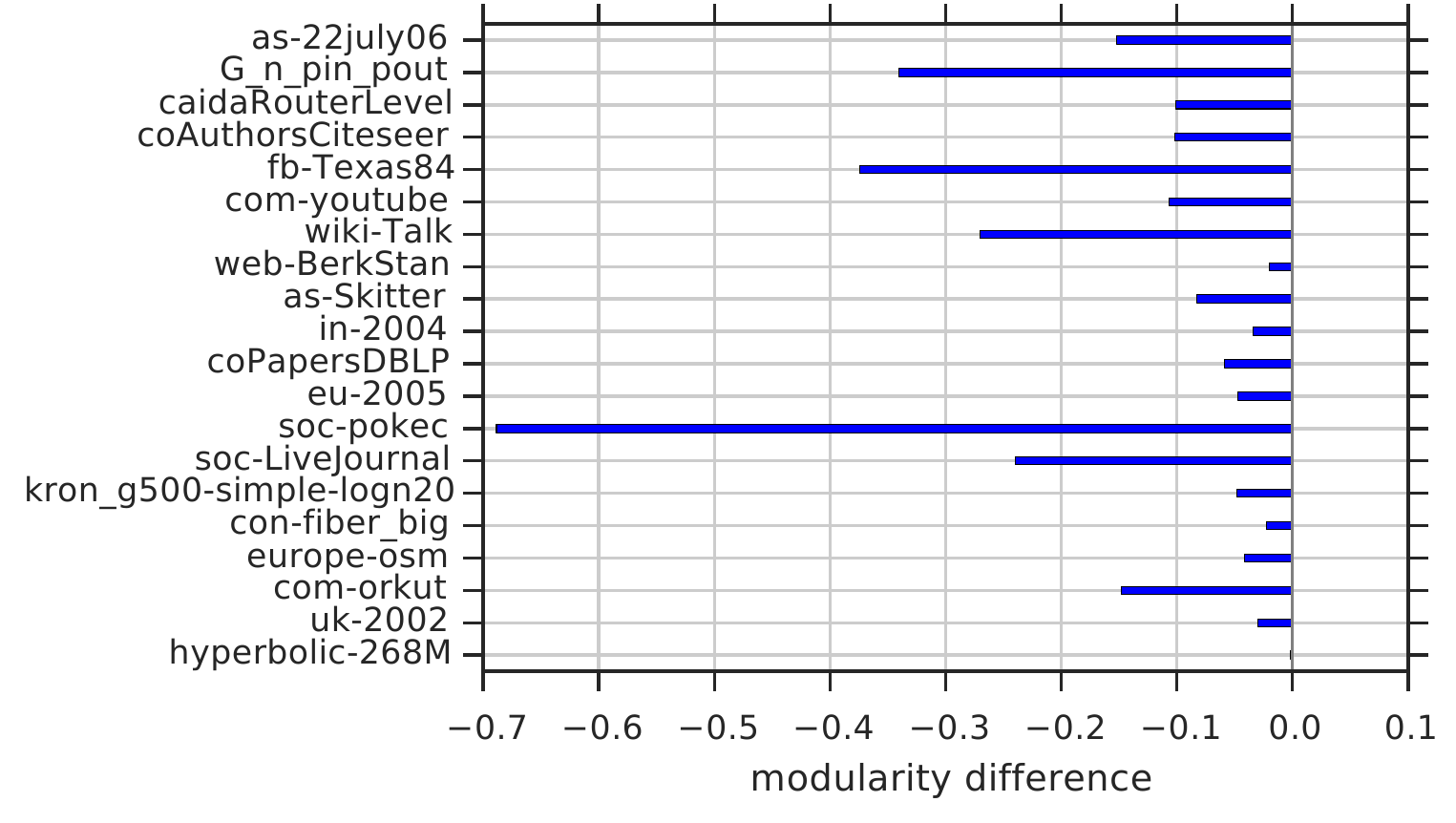}
\includegraphics[width=.45\columnwidth]{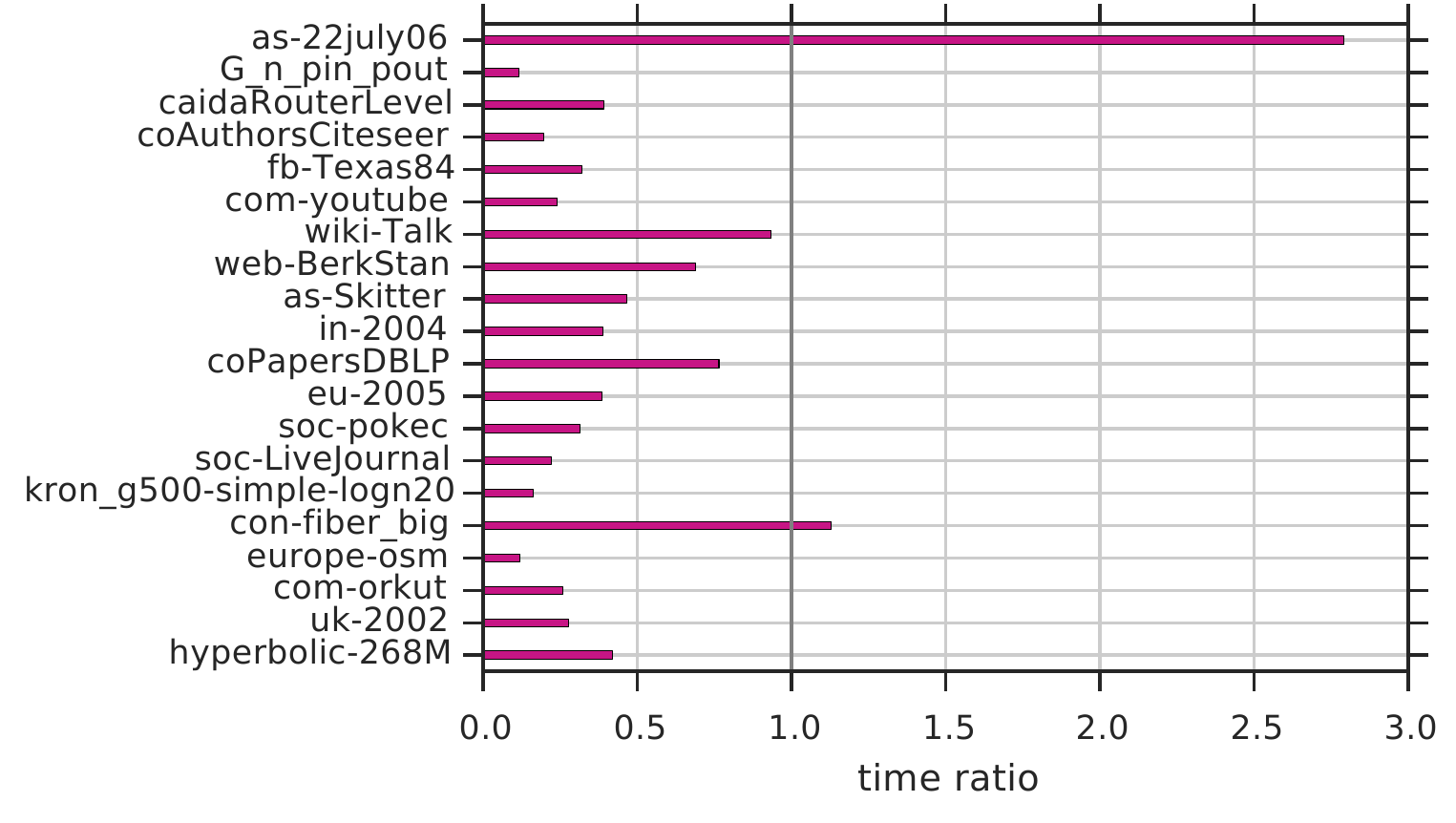}
    }
    \hfill
    \subfloat[\textsf{PLMR} \label{fig:plmr-rel}]{%
\includegraphics[width=.45\columnwidth]{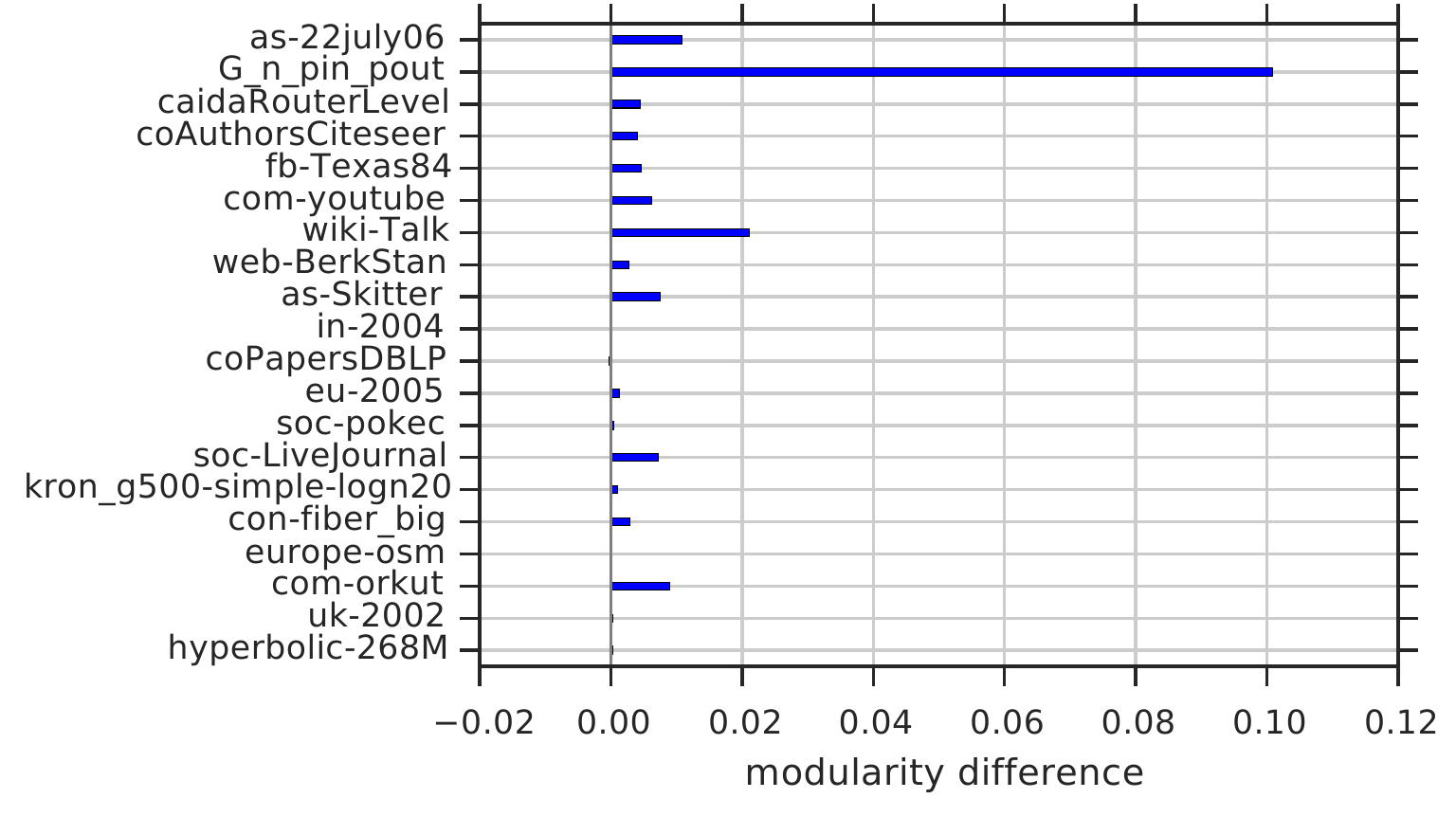}
\includegraphics[width=.45\columnwidth]{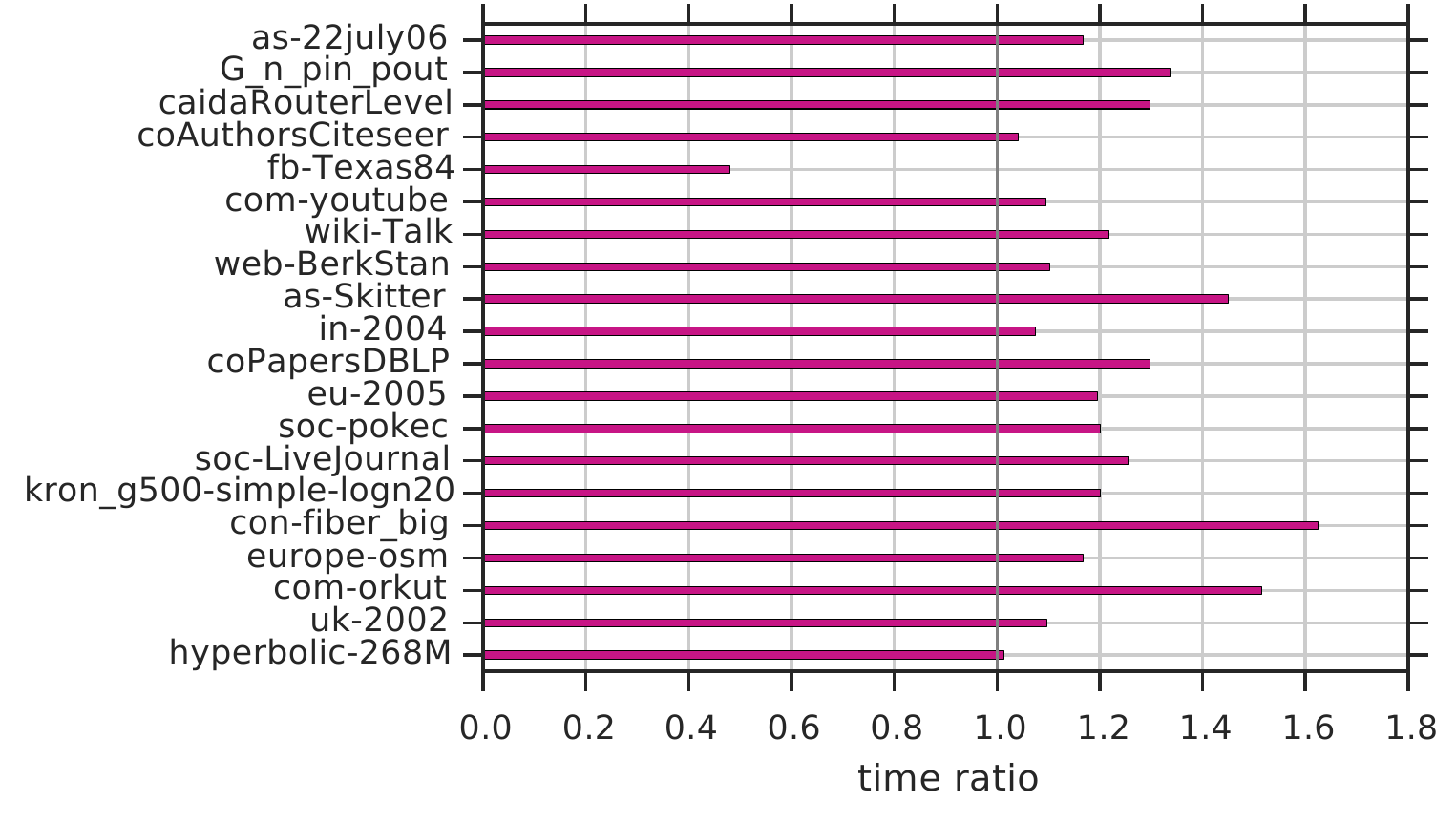}
    }
        \hfill
    \subfloat[\textsf{EPP(4, PLP, PLMR)} \label{fig:epp-plmr-rel}]{%
\includegraphics[width=.45\columnwidth]{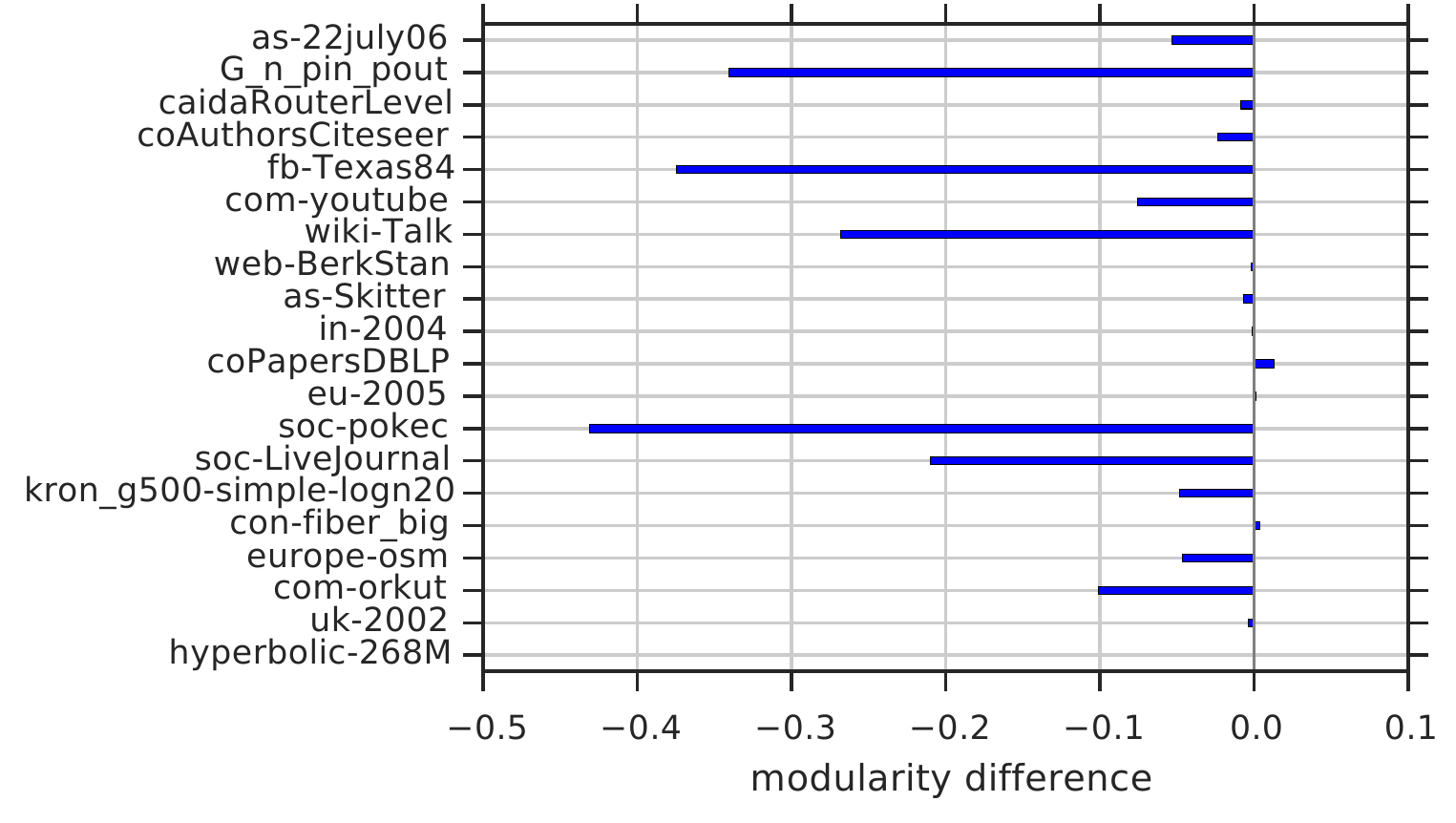}
\includegraphics[width=.45\columnwidth]{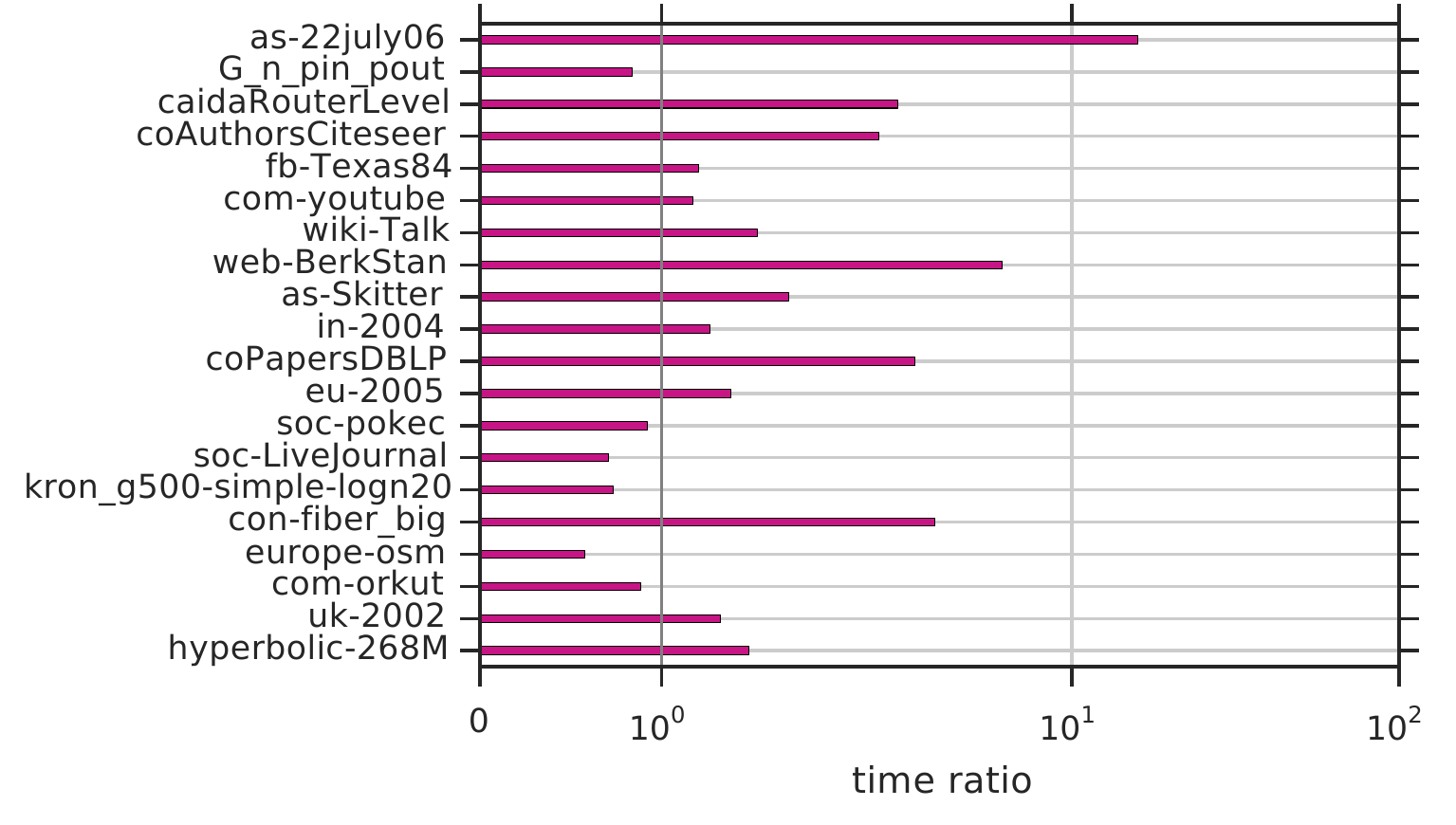}
    }

    \caption{Performance of our algorithms in comparison: \textsf{PLM} serves as the baseline. 32 threads used.}
    \label{fig:performance-ours}
    \end{center}
  \end{figure}

For \textsf{PLM} we observe only small deviations
in quality between single-threaded and multi-threaded runs,
supporting the argument that the algorithm is able to correct undesirable
decisions due to stale data. 
\textsf{PLM} detects communities with relatively high modularity in the majority of networks. 
Even large instances are processed in no more than a few minutes.
Figure~\ref{fig:plm-scaling} shows the scaling behavior of \textsf{PLM}.
Since both the node move phase and the coarsening phase have been parallelized, \textsf{PLM} profits
from increased parallelism as well, achieving a speedup of factor 9 for 32 threads.
In comparison to \textsf{PLP} (Figure~\ref{fig:plp-rel}), we observe that \textsf{PLP} can solve instances 
in only half the time required by \textsf{PLM}, but at a significant loss of modularity.
As discussed in Sec.~\ref{sec:qualitative}, the communities detected by the two algorithms
can be markedly different. 
Because the Louvain method for community detection is well-known and accepted, 
we choose the performance of \textsf{PLM} as our baseline (Figure~\ref{fig:plm-baseline}) and present quality and running time of other algorithms relative to \textsf{PLM}.

\subsection{Parallel Louvain Method with Refinement (\textsf{PLMR})\label{sub:PLMR-Results}}

\begin{figure}[h]
\begin{center}
\includegraphics[width=0.75\columnwidth]{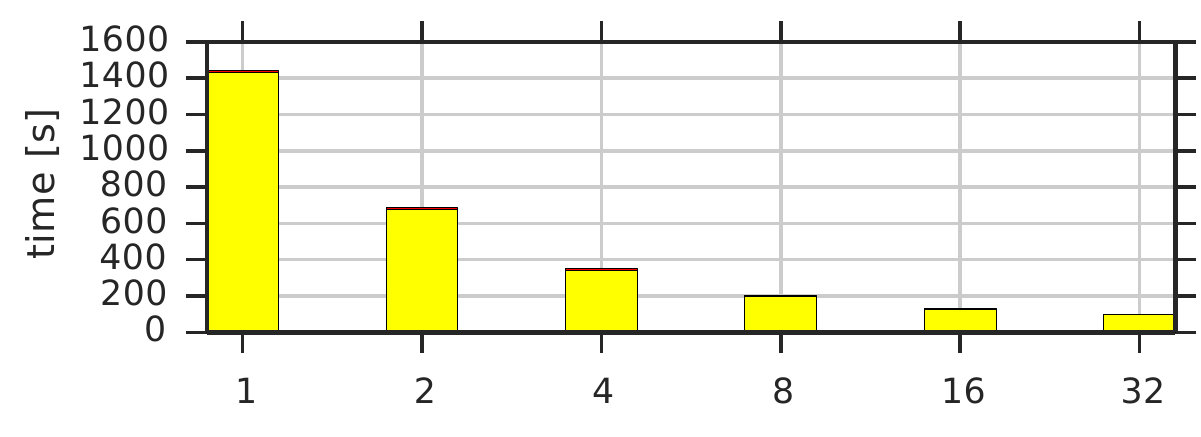}
\includegraphics[width=0.75\columnwidth]{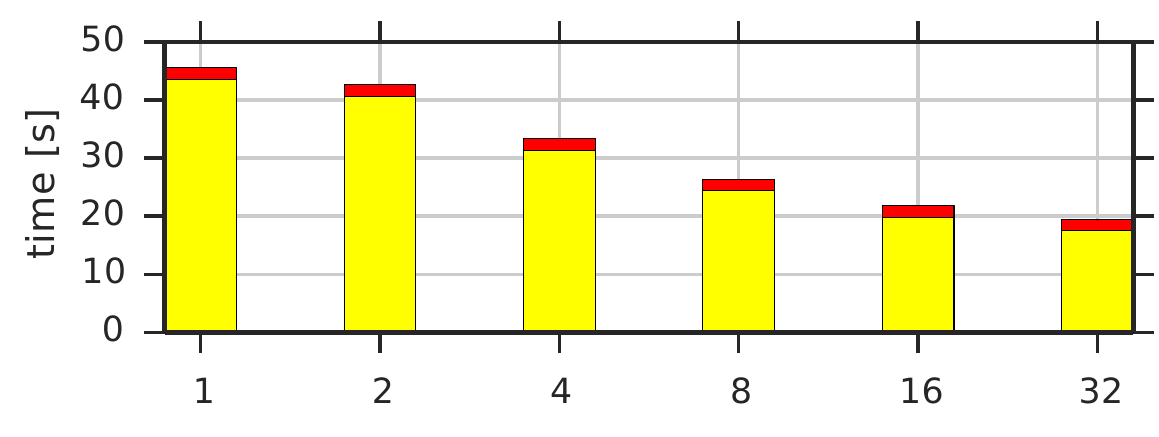}
\includegraphics[width=0.75\columnwidth]{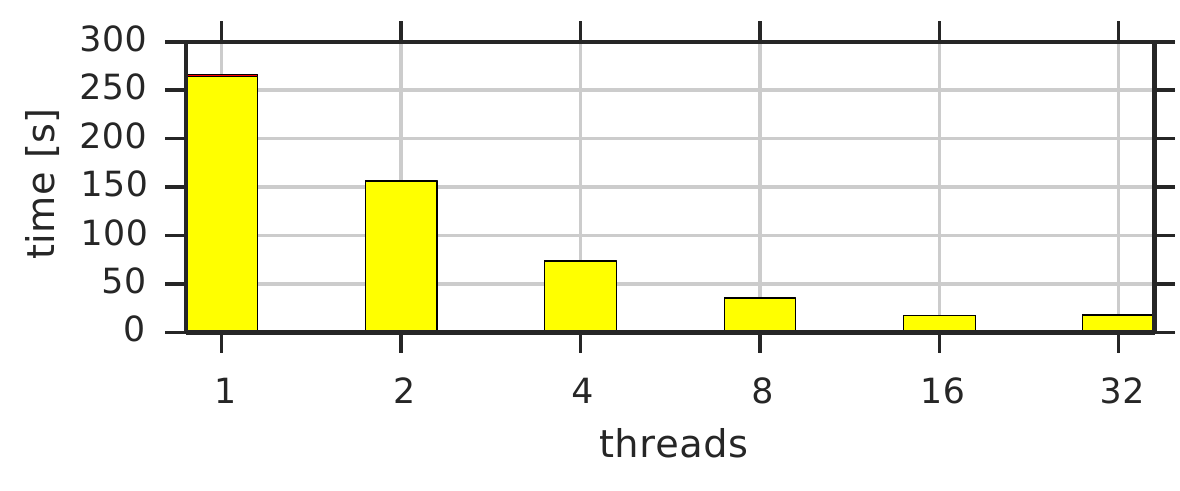}

\caption{\textsf{\label{fig:plm-phases}PLMR} strong scaling of the move, coarsening and refinement phases (top to bottom) on  \texttt{uk-2007-05}}
\end{center}
\end{figure}

\begin{figure}[h]
\begin{center}
\includegraphics[width=0.75\columnwidth]{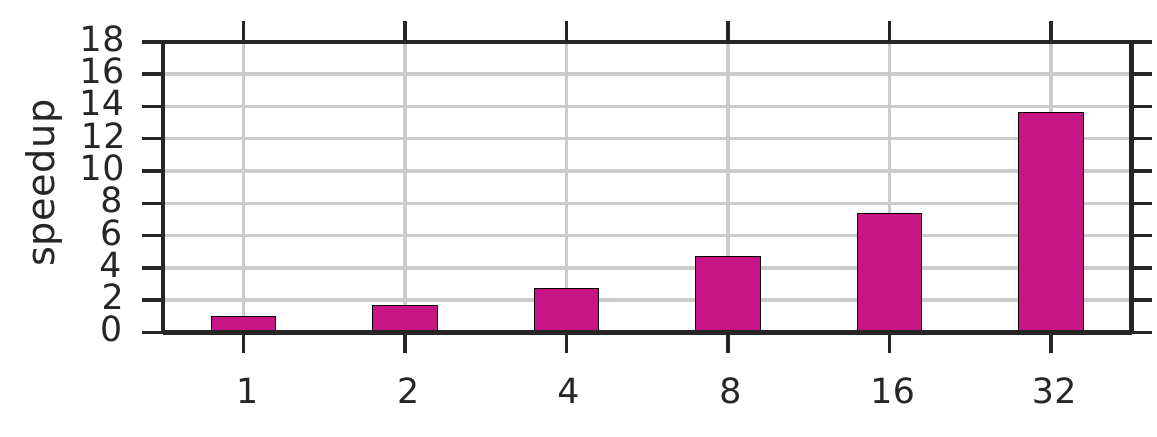}
\includegraphics[width=0.75\columnwidth]{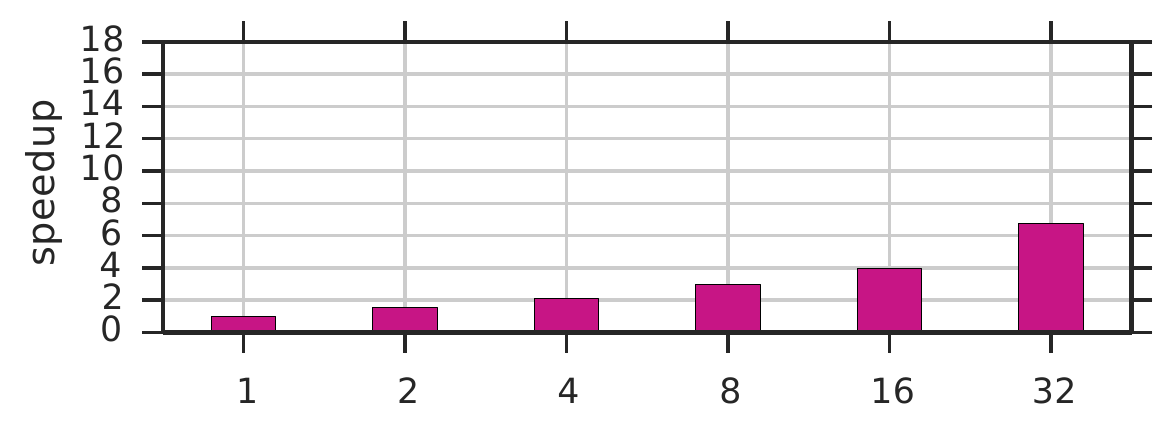}
\includegraphics[width=0.75\columnwidth]{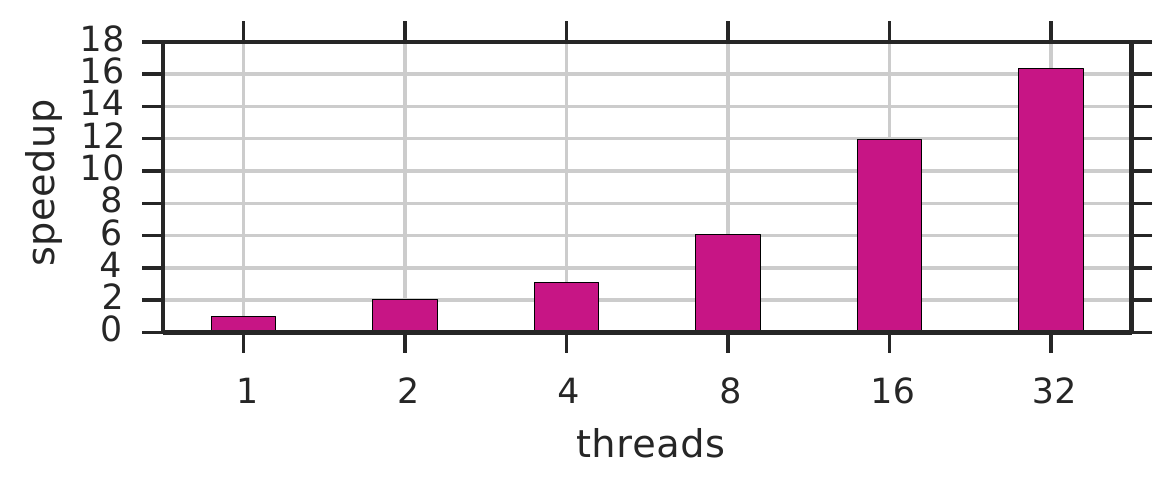}
\caption{\textsf{\label{fig:aggregate-phases}PLMR} strong scaling of the move, coarsening and refinement phases (top to bottom) -- speedup factors aggregated}
\end{center}
\end{figure}

As shown by Figure~\ref{fig:plmr-rel}, adding a refinement phase generally leads to a (sometimes significant) improvement in modularity. This improvement is paid for by a small increase in running time. 
The results indicate that our proposed extension of the original Louvain method by a refinement
phase can efficiently increase solution quality.
We also evaluate the scaling behavior of each phase of the \textsf{PLMR} algorithm.
In Figure~\ref{fig:plm-phases} a yellow bar indicates the running time on the finest graph while the red bar stops at the total running time of the phase. Time spent on the finest graph clearly dominates all running times.
Our experiments show that the move and refinement phases scale well with the number of threads, while the coarsening phase only partially profits from parallelization.
The results on this graph are representative for the trend of the scaling behavior for the algorithm's phases: Figure~\ref{fig:aggregate-phases} shows speedup factors for each of the phases, aggregated over the test set of 20 graphs.

\vspace{-5ex}

\subsection{Ensemble Preprocessing (\textsf{EPP})}

Figure~\ref{fig:EPP-vs-PLP} in the supplementary material demonstrates the effectiveness of the ensemble approach. 
Results were generated by an \textsf{EPP} instance with
a 4-piece \textsf{PLP} ensemble and \textsf{PLMR} as final algorithm
in comparison to a single \textsf{PLP} instance. We observe that the
approach of \textsf{EPP} pays off in the form of improved modularity 
on most instances, exploiting differences in the base solutions and 
spending extra time on classifying contested nodes.
For larger networks, this comes at a cost of about 5 times the running time
of \textsf{PLP} alone.
It also becomes clear that for small networks the approach does not pay off as
running time becomes dominated by the overhead of the ensemble scheme.
In comparison to \textsf{PLM} (Figure~\ref{fig:epp-plmr-rel}), the ensemble approach can be slightly
faster on some networks, but quality is slightly worse in most cases.
We conclude that the ensemble technique \textsf{EPP} is effective
in improving on the quality of a single algorithm. While somewhat lower in modularity,
the communities detected are similar (see Sec.~\ref{sec:qualitative}) to 
those of the Louvain method. 
In practice, our acceleration of the \textsf{PLM} algorithm have made the
ensemble approach less relevant.

\subsection{Comparison with State-of-the-Art Competitors\label{sub:Comparison}}

In this section we present results for an experimental comparison with 
several relevant competing community detection codes.
These are mainly those which excelled in the\emph{ DIMACS} challenge
either by solution quality or time to solution: The 
agglomerative algorithms \textsf{CLU\_TBB}\footnote{\textsc{CLU\_TBB} \url{http://www.staff.science.uu.nl/~faggi101/}} and
\textsf{RG}, as well as \textsf{CGGC} and \textsf{CGGCi}\footnote{\textsc{RG} etc: \url{http://www.umiacs.umd.edu/~mov/}}, ensemble algorithms based on \textsf{RG}.
We also include the widely used original sequential \textsf{Louvain}\footnote{\textsf{Louvain} \url{https://sites.google.com/site/findcommunities/}} implementation, as well as the agglomerative algorithm \textsf{CEL}.
In contrast to the \textit{DIMACS} challenge,
we run all codes on the same multicore machine (Tab.~\ref{tab:Platform}) and 
measure time to solution for sequential and parallel ones alike.

\paragraph{\textsf{Louvain}}

\begin{figure}[!p]
  \begin{center}
        \subfloat[\textsf{Louvain} \label{fig:louvain-rel}]{%
\includegraphics[width=.45\columnwidth]{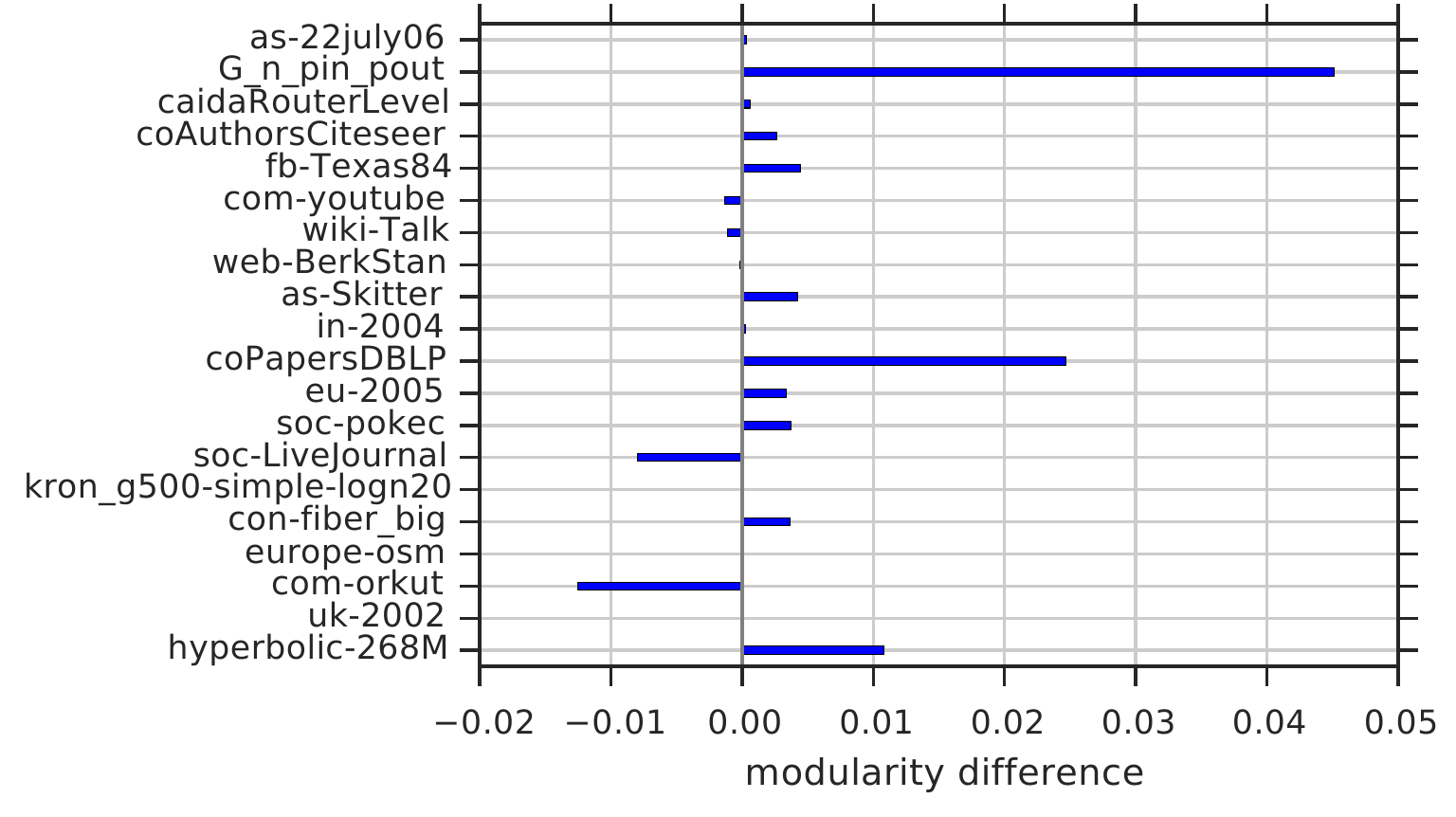}
\includegraphics[width=.45\columnwidth]{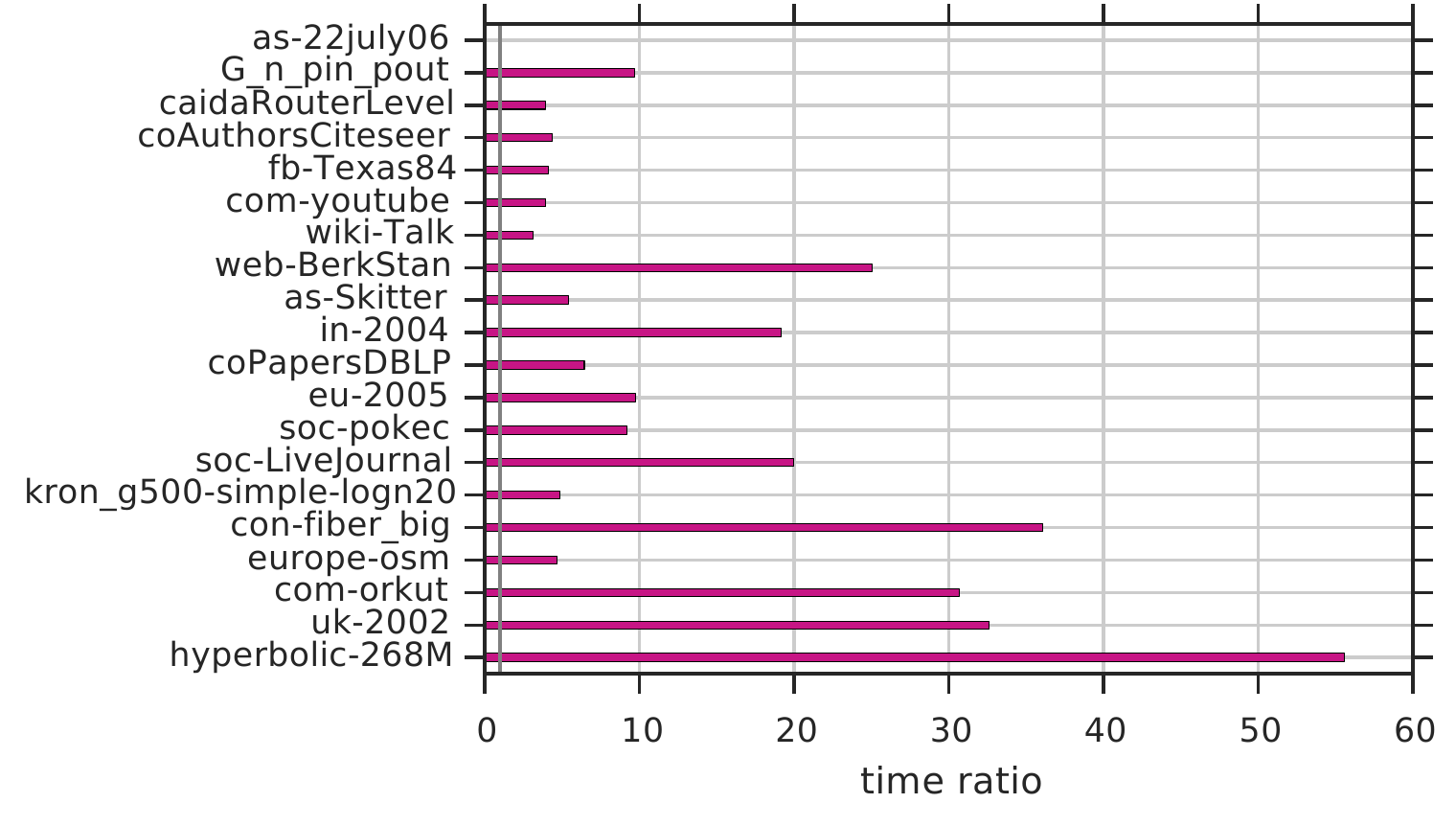}
    }
    \hfill
      \subfloat[\textsf{CLU\_TBB} \label{fig:clutbb-rel}]{%
\includegraphics[width=.45\columnwidth]{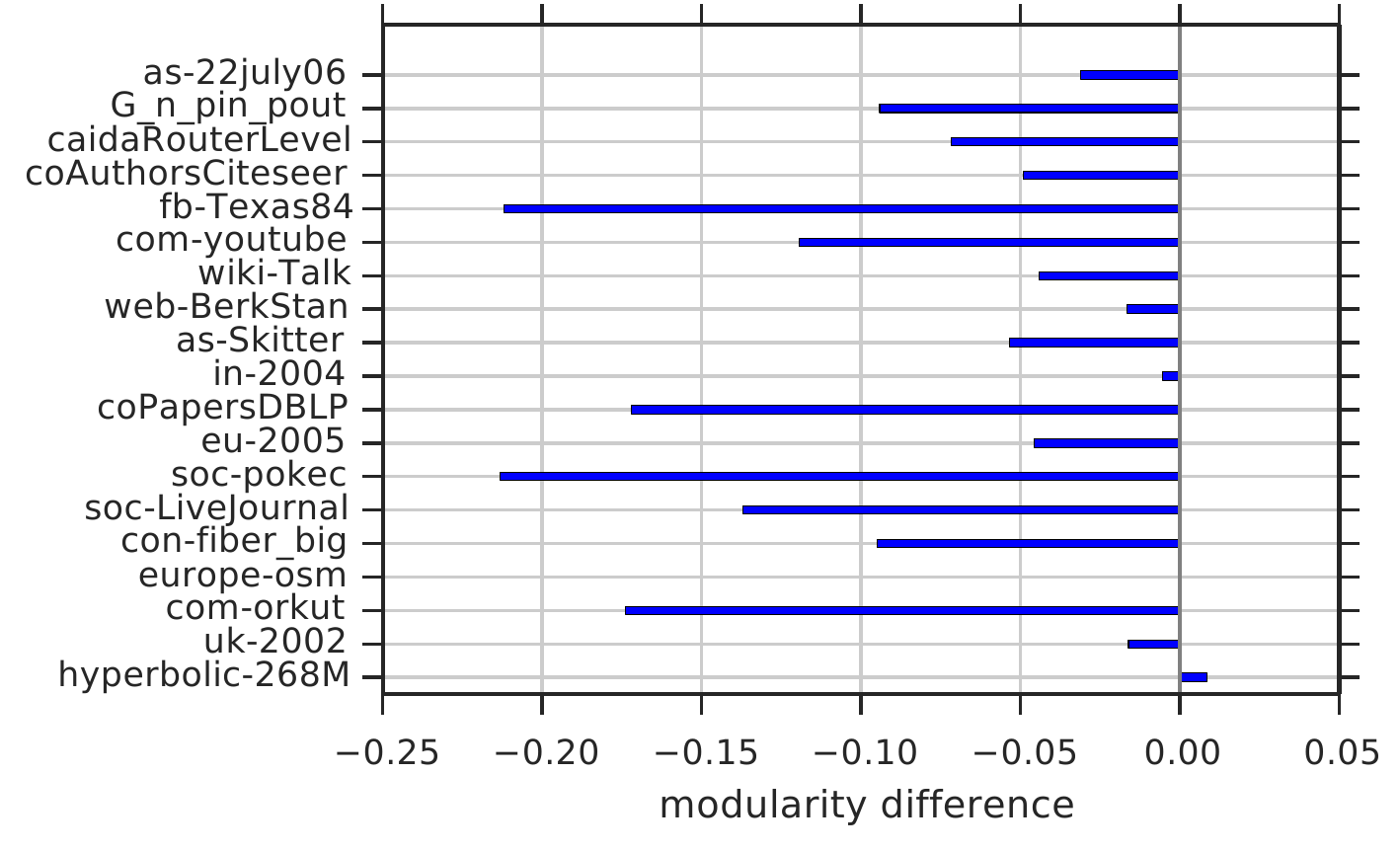}
\includegraphics[width=.45\columnwidth]{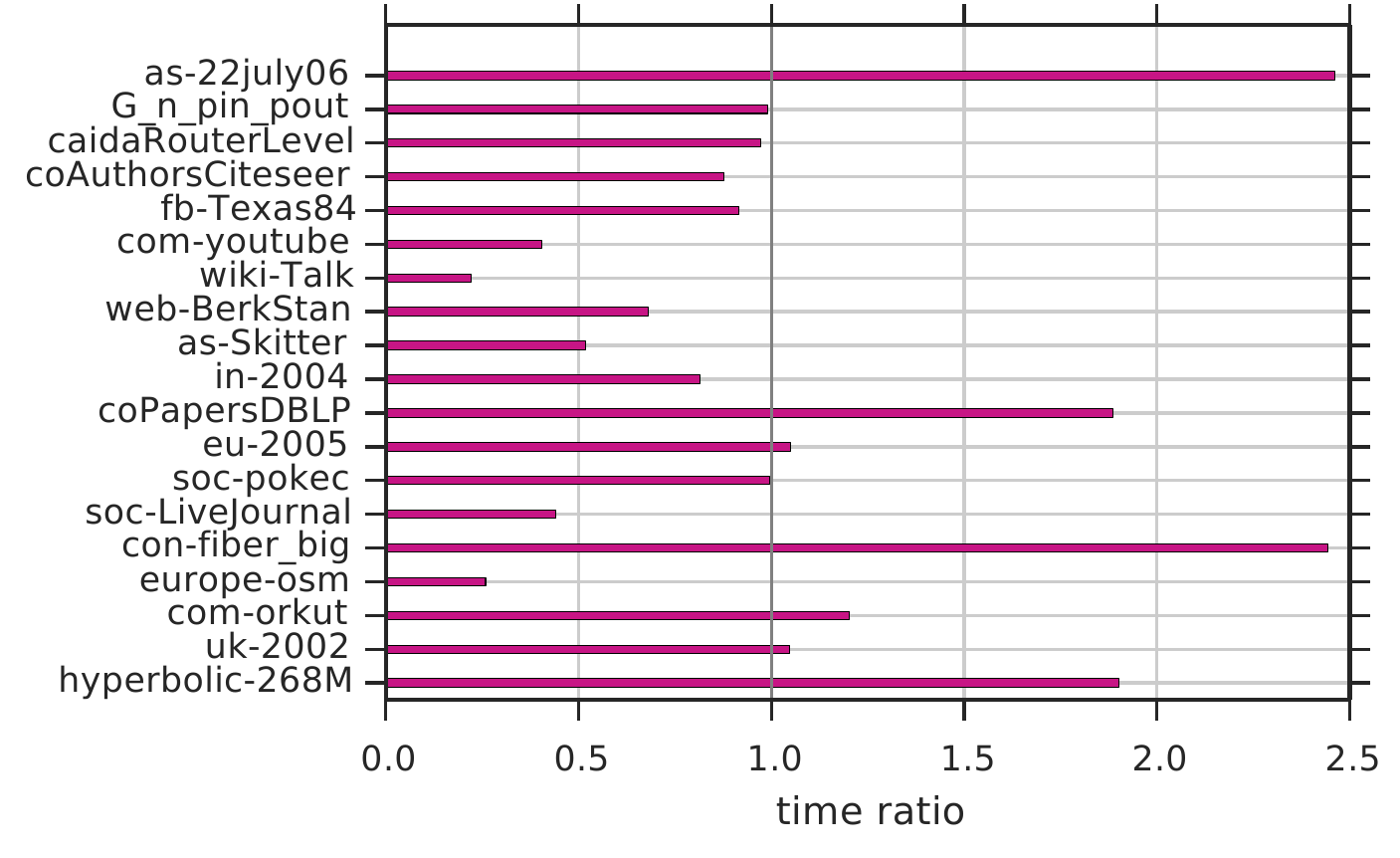}
    }
    \hfill
          \subfloat[\textsf{RG} \label{fig:rg-rel}]{%
\includegraphics[width=.45\columnwidth]{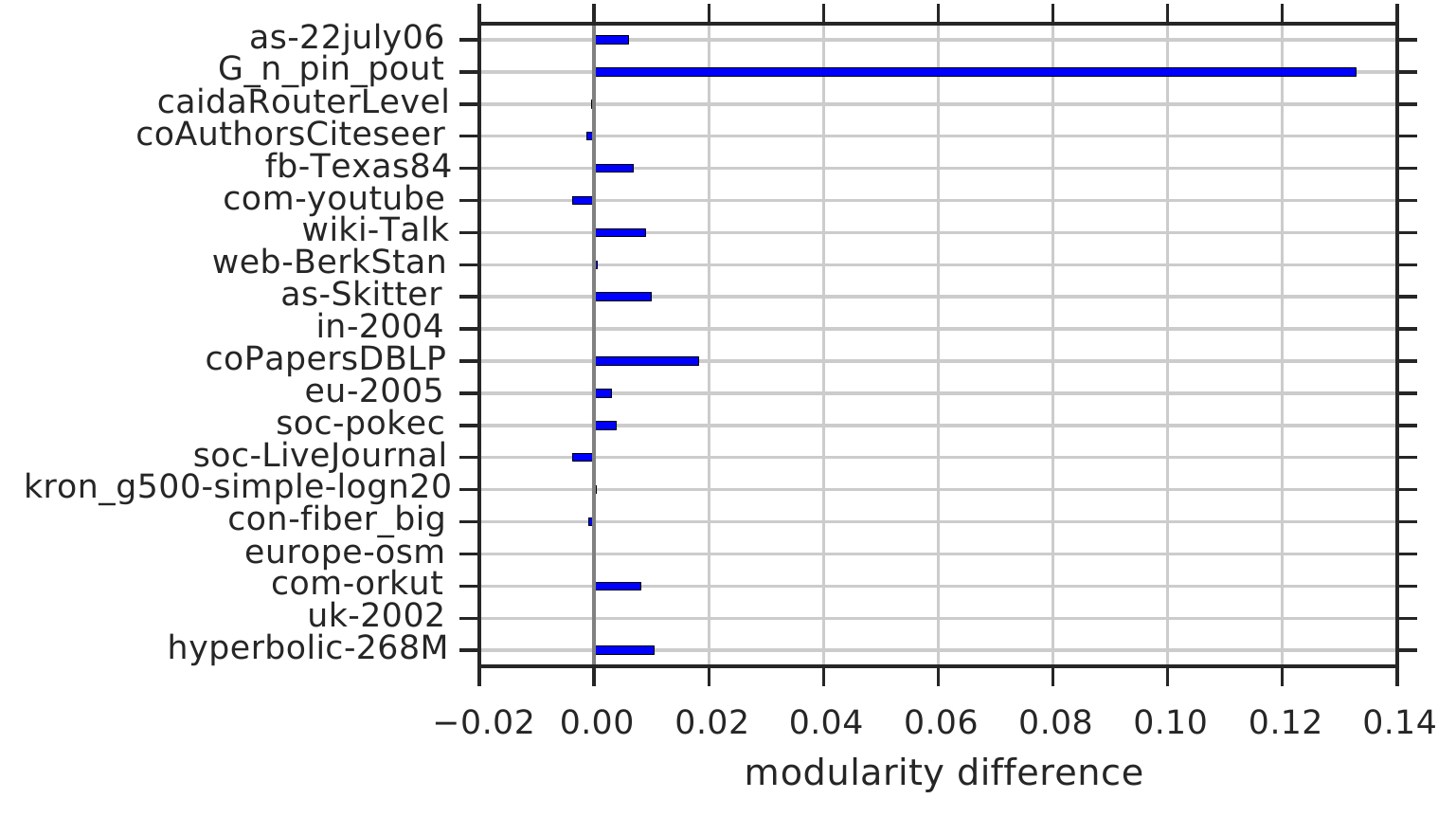}
\includegraphics[width=.45\columnwidth]{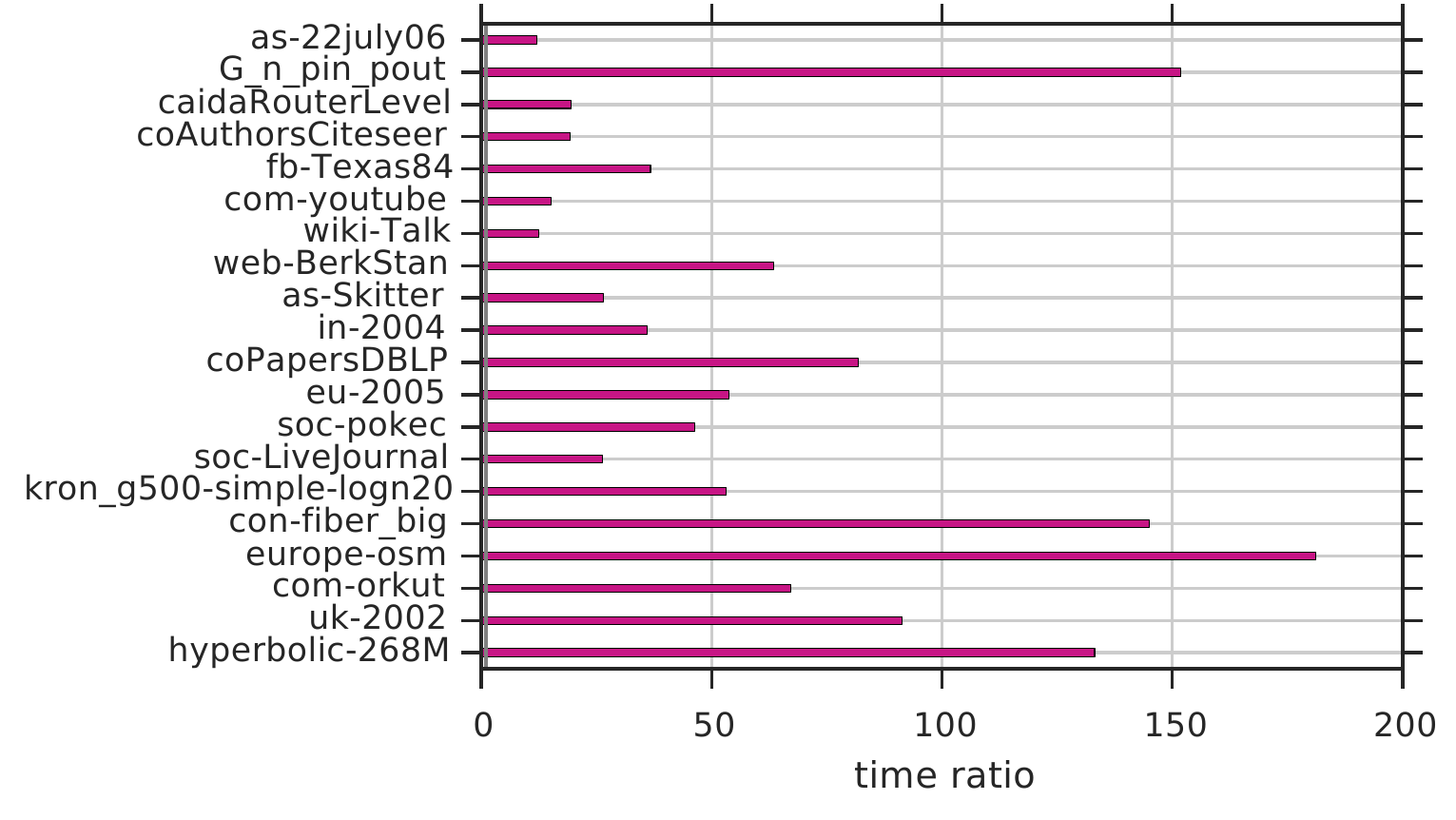}
    }
   \hfill
\subfloat[\textsf{CGGC} \label{fig:cggc-rel}]{%
\includegraphics[width=.45\columnwidth]{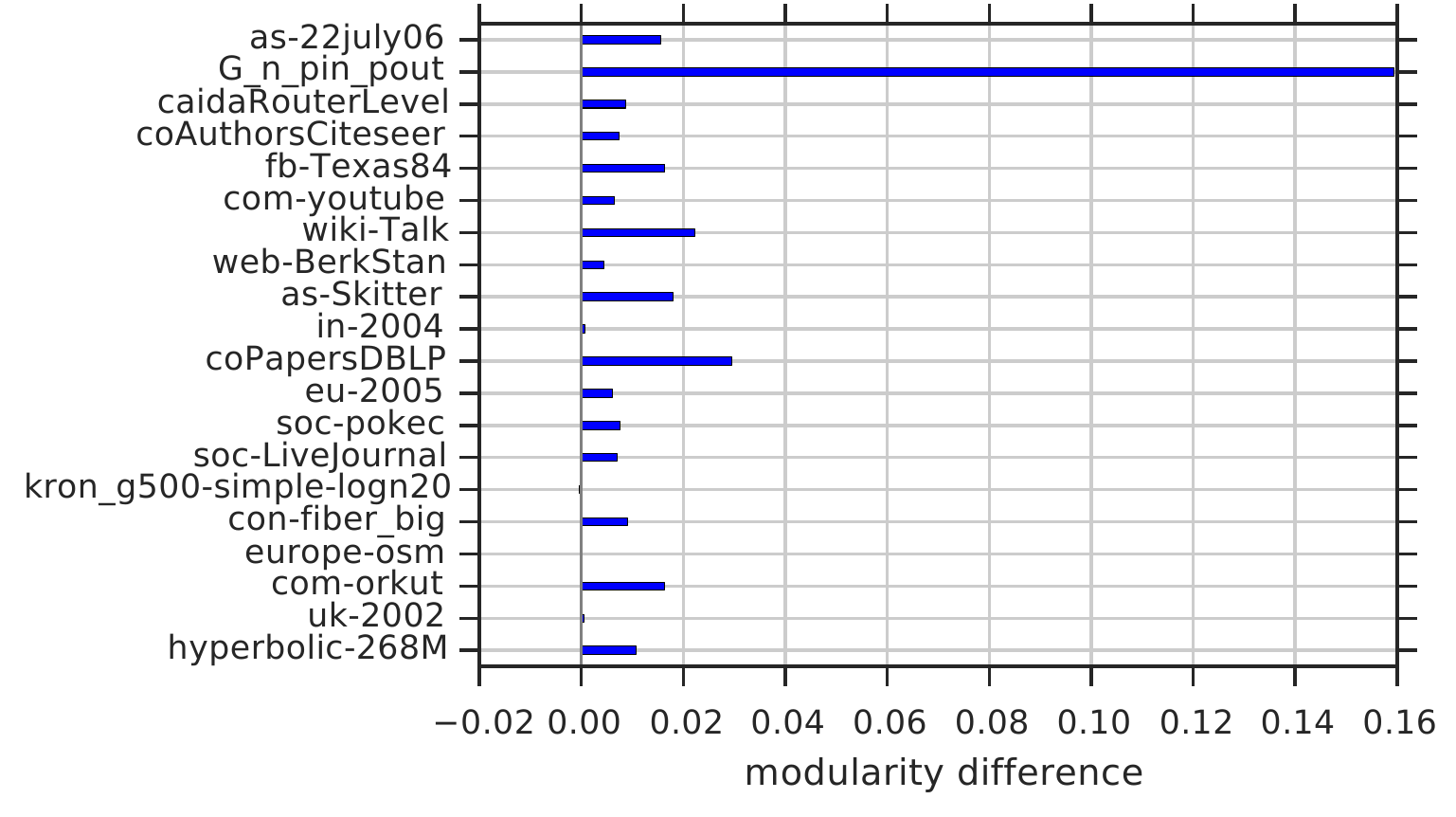}
\includegraphics[width=.45\columnwidth]{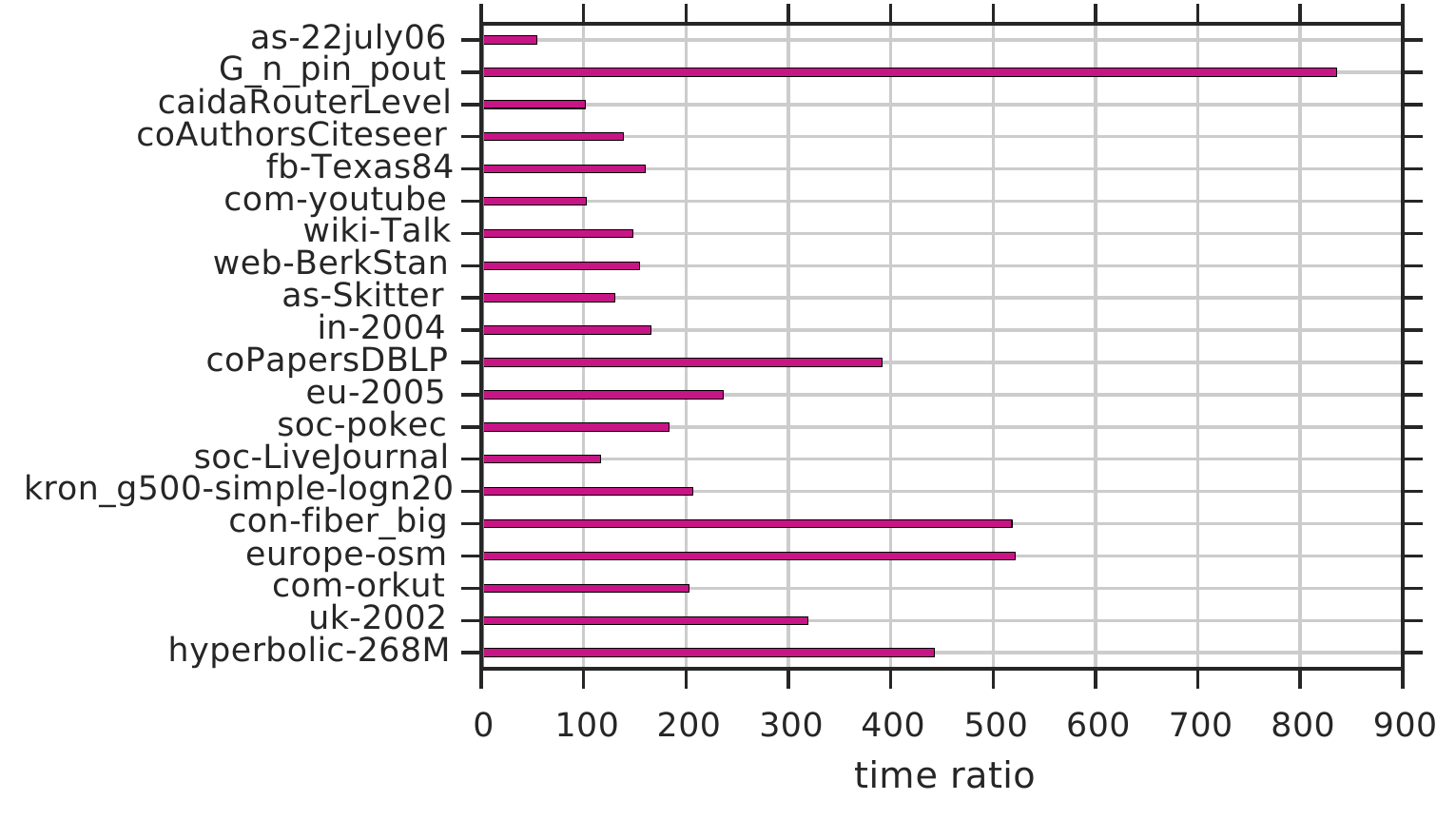}
    }
   \hfill
    \caption{Performance of competitors relative to baseline \textsf{PLM}. 32 threads used for CLU\_TBB.}
    \label{fig:performance-theirs}
    \end{center}
  \end{figure}

Although not submitted to the \textit{DIMACS} competition, the original sequential implementation of the Louvain method 
is still relatively fast (Figure~\ref{fig:louvain-rel}).
The marginally different modularity values in comparison to \textsf{PLM} may be caused by subtle differences in the implementation. For example, \textsf{Louvain} explicitly randomizes the order in which nodes are visited, while we rely on implicit randomization through parallelism. For the smallest graphs, running time values are missing because the implementation reported a running time of zero. \textsf{Louvain} eventually falls behind the parallel algorithm for large graphs, confirming that the overhead and complexity introduced by parallelism is eventually justified when we target massive datasets. 

\paragraph{\textsf{CLU\_TBB} and \textsf{CEL}}

\textsf{CLU\_TBB}, one of the few parallel entries in the \textit{DIMACS} competition, is a very fast implementation of agglomerative modularity maximization, solving the larger instances more quickly than \textsf{PLM} (Figure~\ref{fig:clutbb-rel}). Qualitatively however, \textsf{PLM} is clearly superior on most networks. Both in terms of modularity and running time, \textsf{CLU\_TBB} occupies a middle ground between \textsf{PLP} and \textsf{PLM}, and is qualitatively very similar to our ensemble algorithm \textsf{EPP}. \textsf{CEL}, as another fast parallel program, produced consistently and significantly worse modularity than \textsf{PLM}, failed to produce a solution on some graphs, and is not as fast as \textsf{PLP}. 

\paragraph{\textsf{RG}, \textsf{CGGC} and \textsf{CGGCi}}

Ovelgönne and Geyer-Schulz entered the DIMACS challenge with an ensemble approach conceptually similar to what we
have developed in this paper. Their base algorithm is the sequential agglomerative \textsf{RG}, and two ensemble variants exist: \textsf{CGGC} implements anensemble technique very similar to \textsf{EPP}, while \textsf{CGGCi} iterates the approach.
The \textsf{RG} algorithm achieves a high solution quality, surpassing \textsf{PLM} by a small margin on most networks (Figure~\ref{fig:rg-rel}). Quality is again slightly improved by the ensemble approach \textsf{CGGC} and its iterated version \textsf{CGGCi} (Figure~\ref{fig:cggc-rel}, and~\ref{fig:cggci-rel} in the SM), with the latter surpassing any other heuristic known to us. However, all three are very expensive in terms of computation time,
often taking orders of magnitude longer than \textsf{PLM}.
We consider running times of several hours for many of our networks no longer viable for the scenario we target, namely interactive data analysis on a parallel workstation.

\subsection{Pareto Evaluation}
\label{sub:Pareto}

\begin{figure}[h]
\begin{center}
\includegraphics[width=.9\columnwidth]{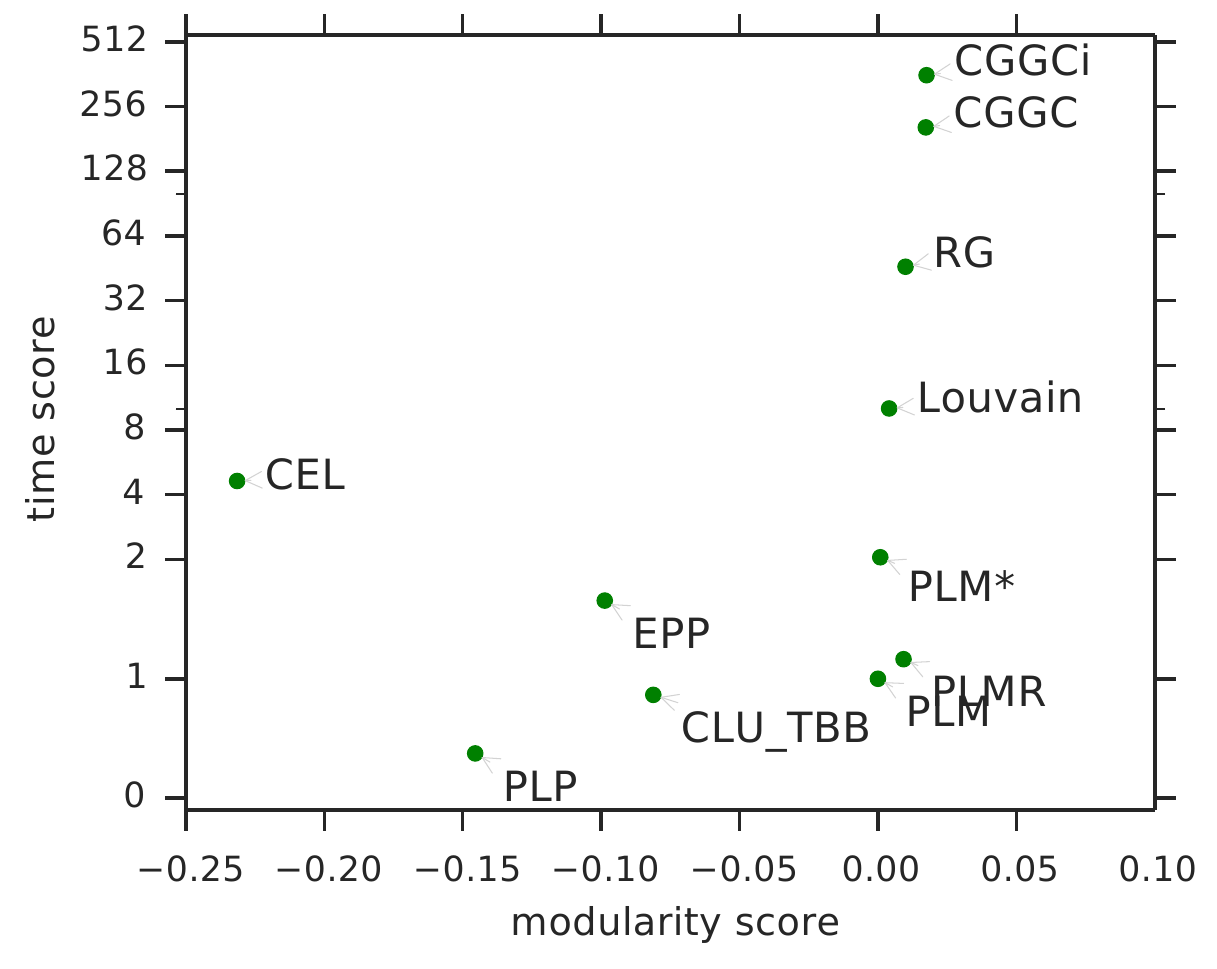}
\caption{Pareto evaluation of community detection algorithms}
\label{fig:pareto}
\end{center}
\end{figure}

We have so far presented results broken down by data set to stress that observed effects may vary strongly from one network to another, a sign of the heterogeneity of real-world complex networks. Additionally, we want to give a condensed picture of the results. For this purpose we use the previous experimental data to compute a score for running time and solution quality. 
The time score is the geometric mean of running time ratios over our test set of networks with the running time
 of \textsc{PLM} as the baseline, while the modularity score is the arithmetic mean of absolute modularity differences. 
Figure~\ref{fig:pareto} shows the resulting points. It becomes clear that all algorithms except \textsf{CEL} and \textsf{EPP} are placed on or close to the Pareto frontier.
\textsf{PLP} is unrivaled in terms of time to solution, but solution quality is suboptimal.
In the middle ground between label propagation and Louvain method, the parallel \textsf{CLU\_TBB} achieves about the same modularity but beats the ensemble approach in terms of speed.
\textsf{PLM} and \textsf{PLMR} emerge as qualitatively strong and fast candidates, closest to the lower right corner.
(Their more memory-efficient implementation \textsf{PLM*} is about a factor of 2 slower.)
It is also evident that our extended version \textsf{PLMR} can improve solution quality for a small computational extra charge. We recommend both \textsf{PLM} and \textsf{PLMR} as the default algorithms for parallel community detection in large networks. The original sequential implementation of the Louvain method is thus no longer on the Pareto frontier since it cannot benefit from multicore systems.
\textsf{RG} and its ensemble combinations have the best modularity scores by a narrow margin, while they are by far the most computationally expensive ones, which places them outside of the application scenario we target.

\vspace{-5ex}

\subsection{LFR Benchmark \label{sub:LFR}}

\begin{figure}[h]
\begin{centering}
\includegraphics[width=\columnwidth]{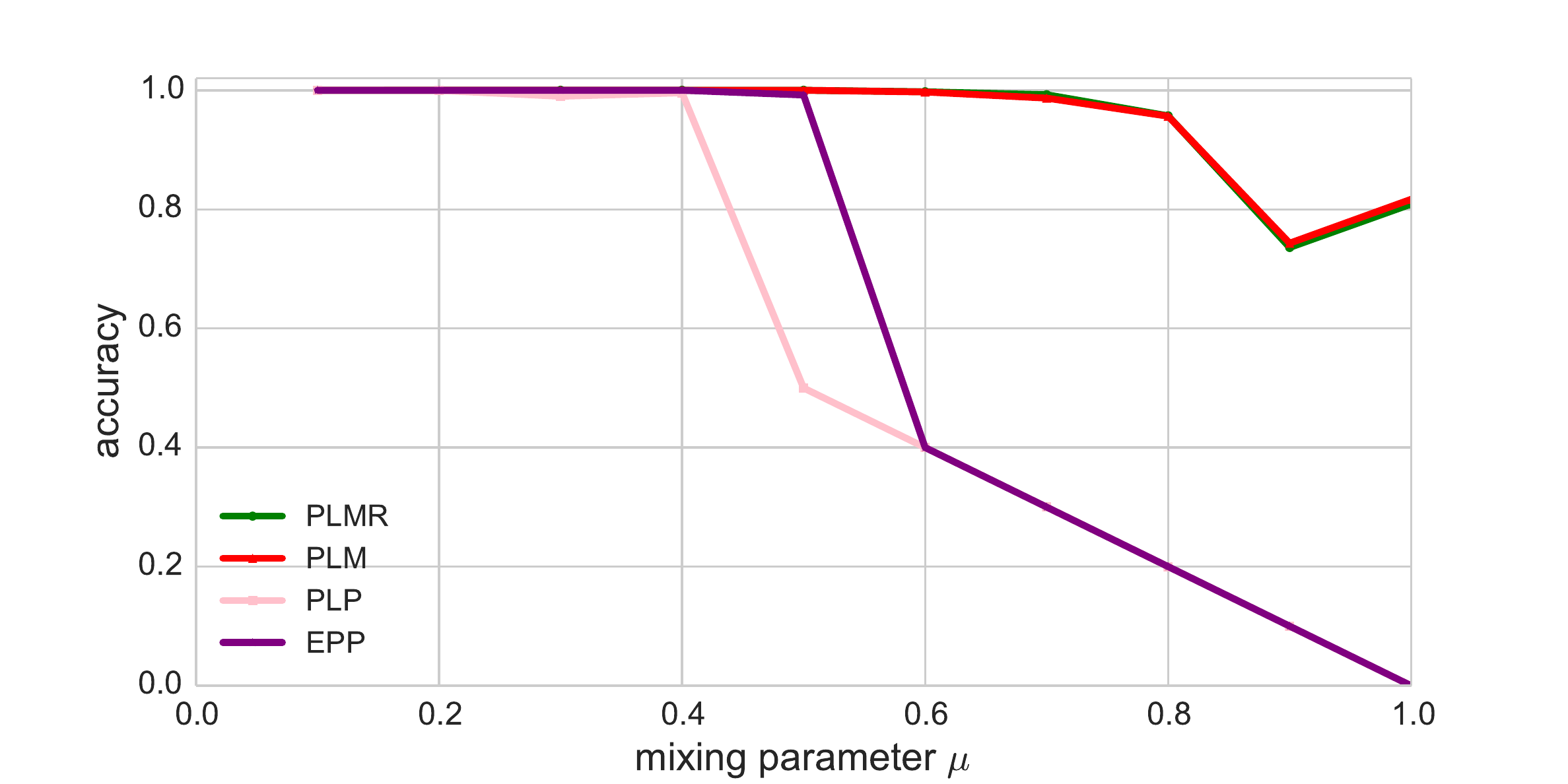}
\par\end{centering}
\caption{LFR benchmark ($n = 10^5$): accuracy in recognizing ground truth while increasing inter-community edges}
\label{fig:LFR}
\end{figure}

The LFR benchmark~\cite{lancichinetti2008benchmark} is an established method for evaluating community detection algorithms: A generator produces graphs that resemble real complex networks and contain dense communities which are the more sparsely connected the lower the mixing parameter $\mu$. Algorithm performance is measured as the accuracy in recognizing the ground truth communities supplied by the generator, in view of increasing difficulty ($\mu$). Although there are real-world networks that come with supposed ground truth communities (\eg interest-based groups of online social networks in the SNAP collection), we consider 
only a synthetic ground truth reliable enough for our purposes. 
In Figure~\ref{fig:LFR} we plot the agreement (graph-structural Rand index, where 1 is complete agreement) between detected and ground truth communities for our algorithms, and show that the \textsf{PLM} method is able to detect the ground truth even with strong noise ($\mu = 0.8$), while \textsf{PLP} (and hence \textsf{EPP}) is somewhat less robust.

\onecolumn

\twocolumn

\subsection{One More Massive Network}
\label{sub:one-more}

\begin{figure}[h]
\includegraphics[width=0.95\columnwidth]{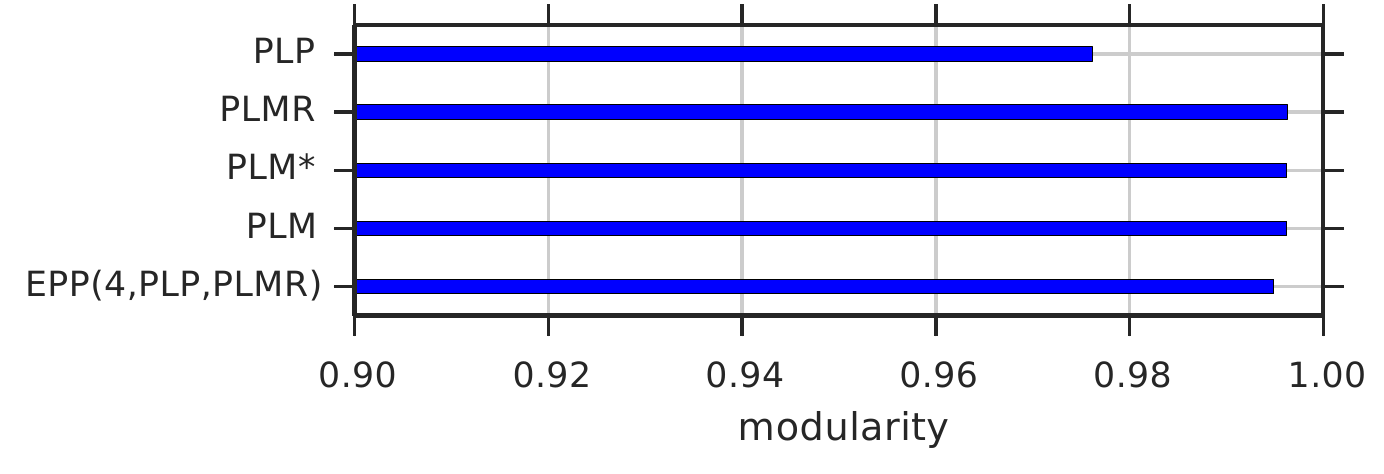}
\includegraphics[width=0.95\columnwidth]{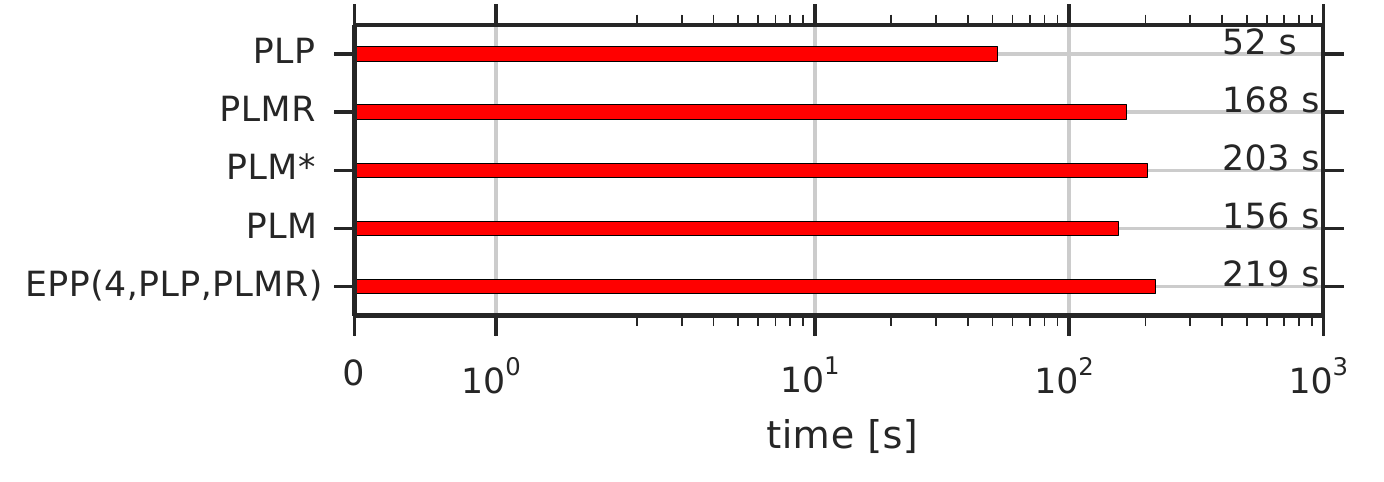}

\caption{\label{fig:uk2007} Modularity and running time at 32 threads for our parallel algorithms on the massive web graph \texttt{uk-2007-05}}
\end{figure}

In addition to the experiments that went into the Pareto evaluation, we run our parallel algorithms on the web graph
 \texttt{uk-2007-05}, at about 3.3 billion edges the largest real-world data set currently available to us.
\textsf{CLU\_TBB} fails at reading the input file. This leaves us with five of our own parallel algorithms for Figure~\ref{fig:uk2007}: \textsf{EPP(4,PLP,PLMR)} takes about 219 seconds, while \textsf{PLM} requires about 156 seconds to arrive at a slightly higher modularity.
As expected, \textsf{PLP} is by far the fastest algorithm and terminates in less than a minute. If a certain modularity loss (here 0.02) is acceptable, \textsf{PLP} is also an appropriate choice for quickly detecting communities in billion-edge networks. The processing rate for \textsf{PLP} is over 53M edges/second and over 21M edges/second for \textsf{PLM} with respect to a complete run of each algorithm. These rates confirm the suitability of our algorithms for analyzing massive complex networks on a commodity shared-memory server.

\subsection{Weak Scaling}

\begin{figure}[h]
\begin{centering}

\includegraphics[height=3.2cm]{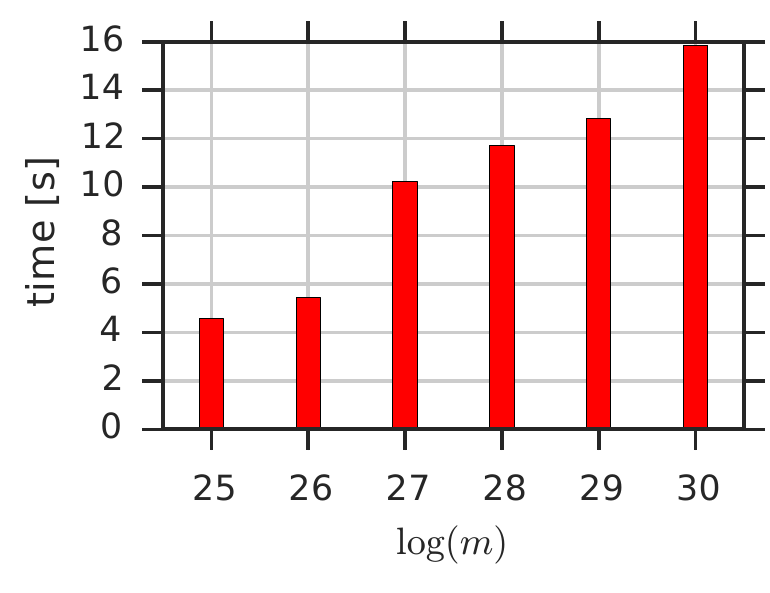}
\includegraphics[height=3.2cm]{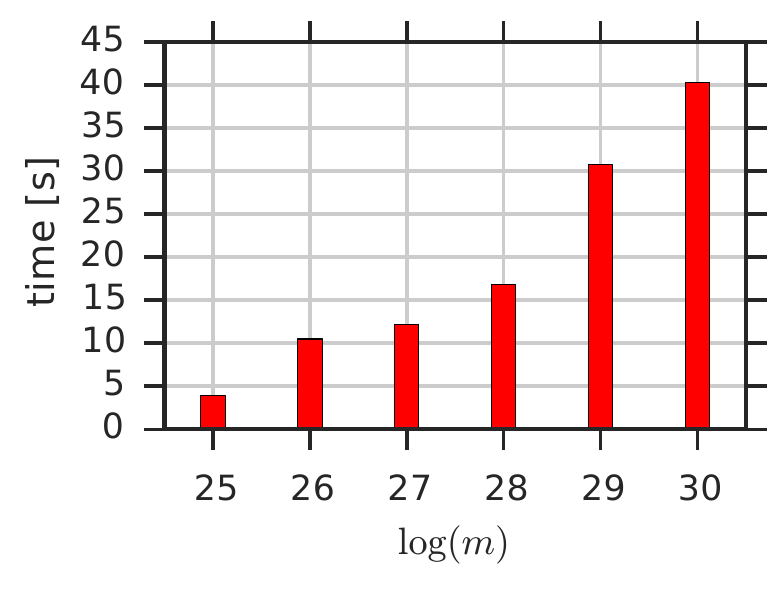}

\caption{\label{fig:plm-weakscaling}\label{fig:plp-weakscaling}
\textsf{PLP} (left) and \textsf{PLM} (right)
weak scaling on the series of HUD graphs}
\end{centering}
\end{figure}

For weak scaling experiments, we use a series of synthetic
graphs where each graph has twice the size of its predecessor (from
$\log m = 25 \dots 30$), and double the number of threads simultaneously
from 1 to 32. 
The graphs were created using a generator~\cite{LSMP14fast} based on a unit-disk graph model in 
hyperbolic geometry~\cite{Krioukov2010} (HUD),
which produces both a power law degree distribution and distinctive dense communities.
Figure~\ref{fig:plp-weakscaling}
shows the results of weak scaling experiments for \textsf{PLP} and \textsf{PLM}. 
It must be noted that perfect scaling cannot be expected due to the complex structure of the input.
The results of the respective last column have been obtained with hyperthreading, which explains the steeper increase.
Fig.~\ref{fig:weakscaling-rmat} in the supplementary material show results for additional weak scaling experiments on synthetic graphs generated with the R-MAT model.

\vspace{-1ex}

\section{Qualitative Aspects}
\label{sec:qualitative}

In this work we concentrate on achieving a good tradeoff between high modularity,
 a widely accepted quality measure for community detection, and low running time. 
Ideally one should also look for further validation of the detected communities beyond good modularity.
This is a difficult task for several reasons. For most networks, we do not have a reliable ground-truth partition, 
especially because community structure is likely a multi-factorial phenomenon in real networks.
Our task is to uncover the hidden community structure of the network.
In order to know whether we have succeeded
in this data mining task, we would have to check whether the solution helps us to  formulate hypotheses to predict and explain real-world phenomena on the basis of network data.
Whether one solution is more appropriate than another may strongly depend on the domain of the network.
Domain-specific validation of this kind goes beyond the scope of this paper
as we focus on parallelization aspects. Also, most sequential counterparts of our algorithms
have been validated before, see \eg~\cite{Blondel:2008uq}.

\begin{figure}
\begin{center}
\includegraphics[width=.65\columnwidth]{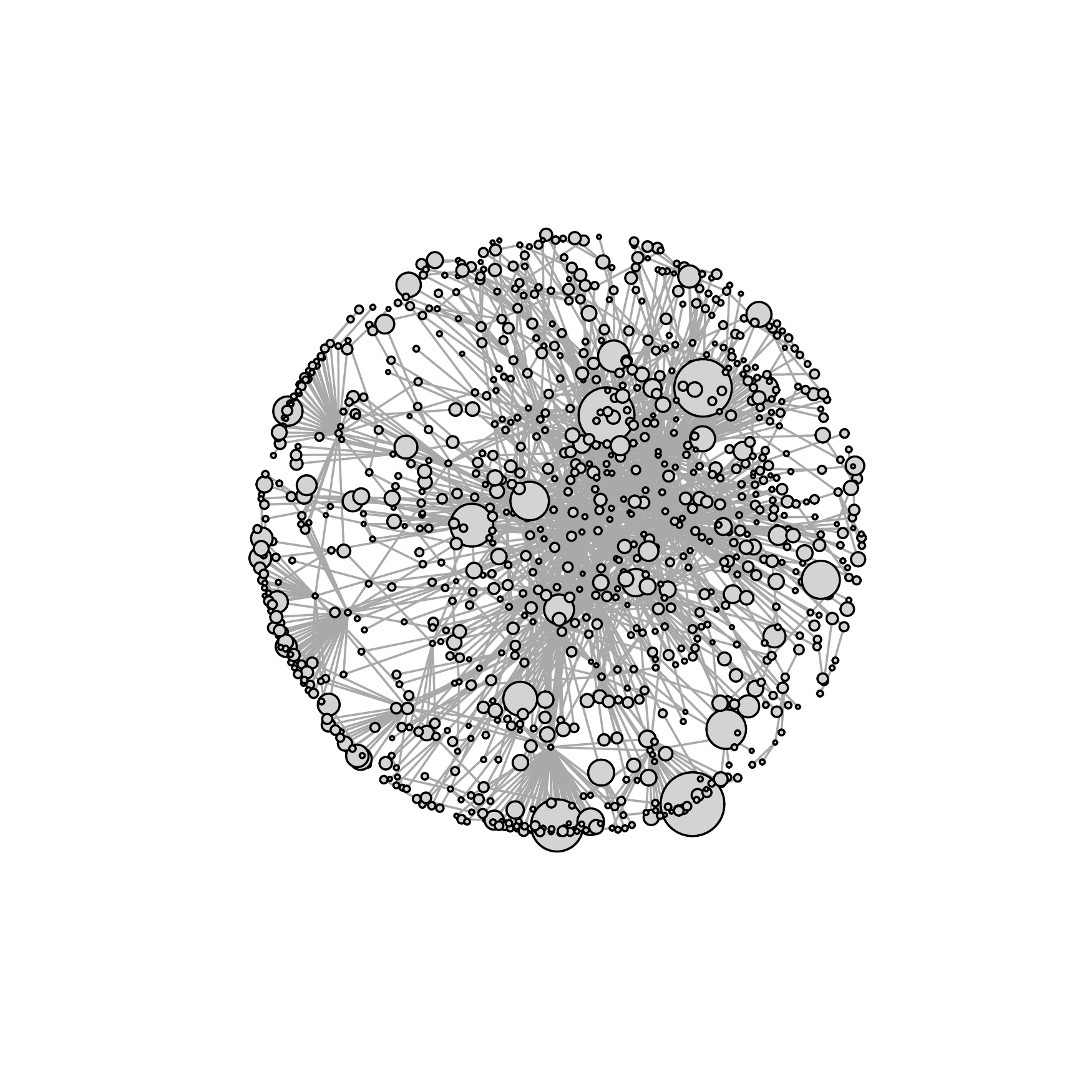}
\includegraphics[width=.65\columnwidth]{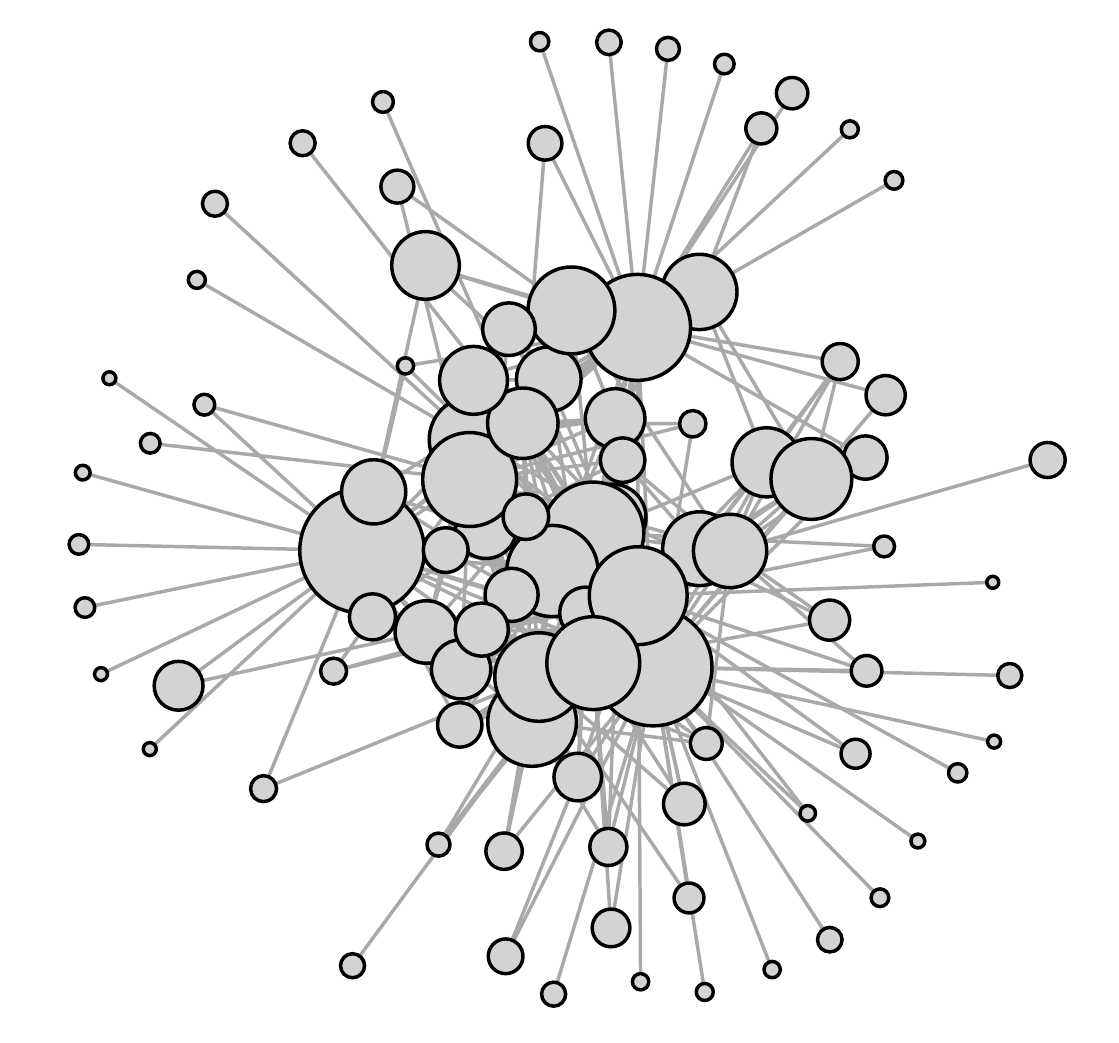}
\includegraphics[width=.65\columnwidth]{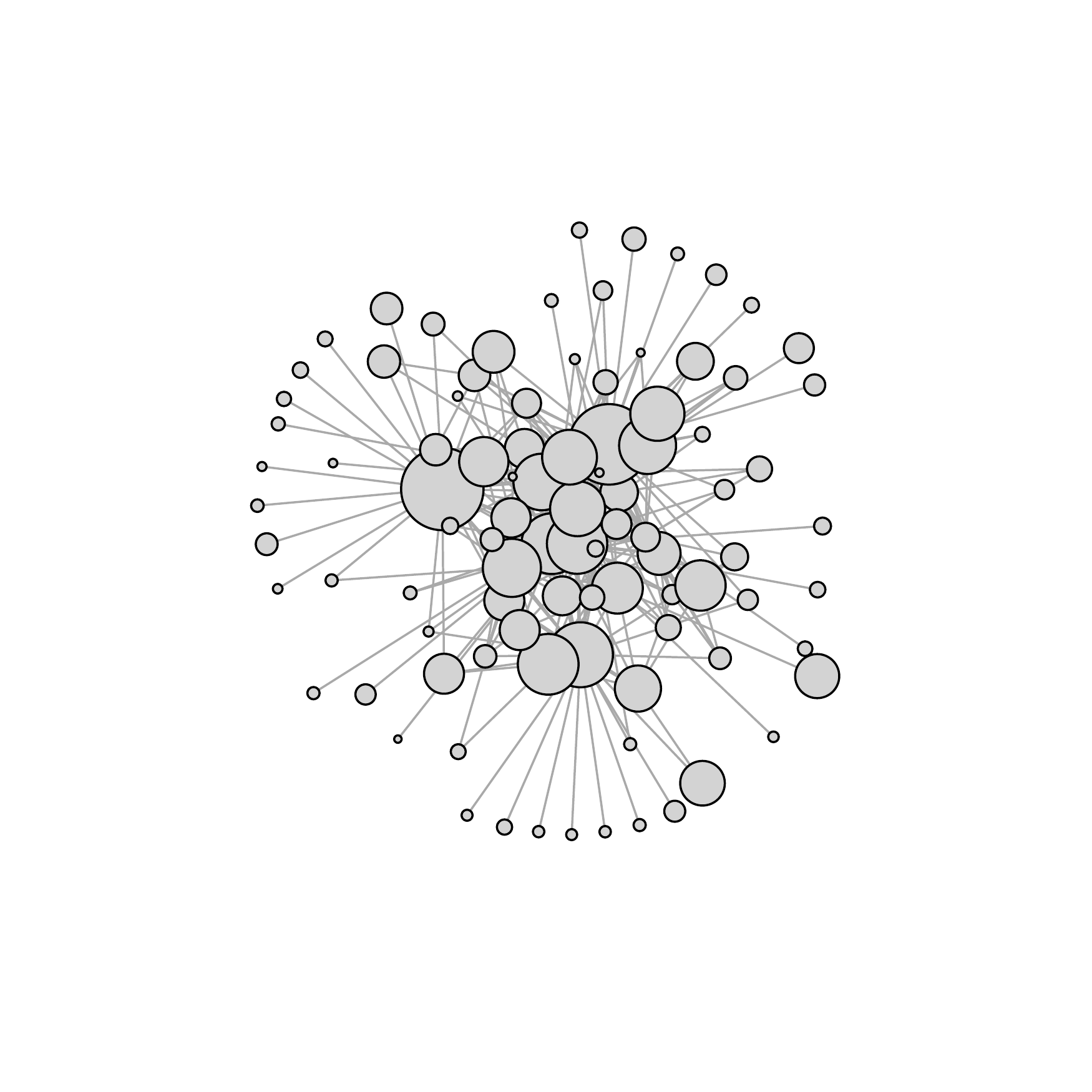}
\includegraphics[width=.65\columnwidth]{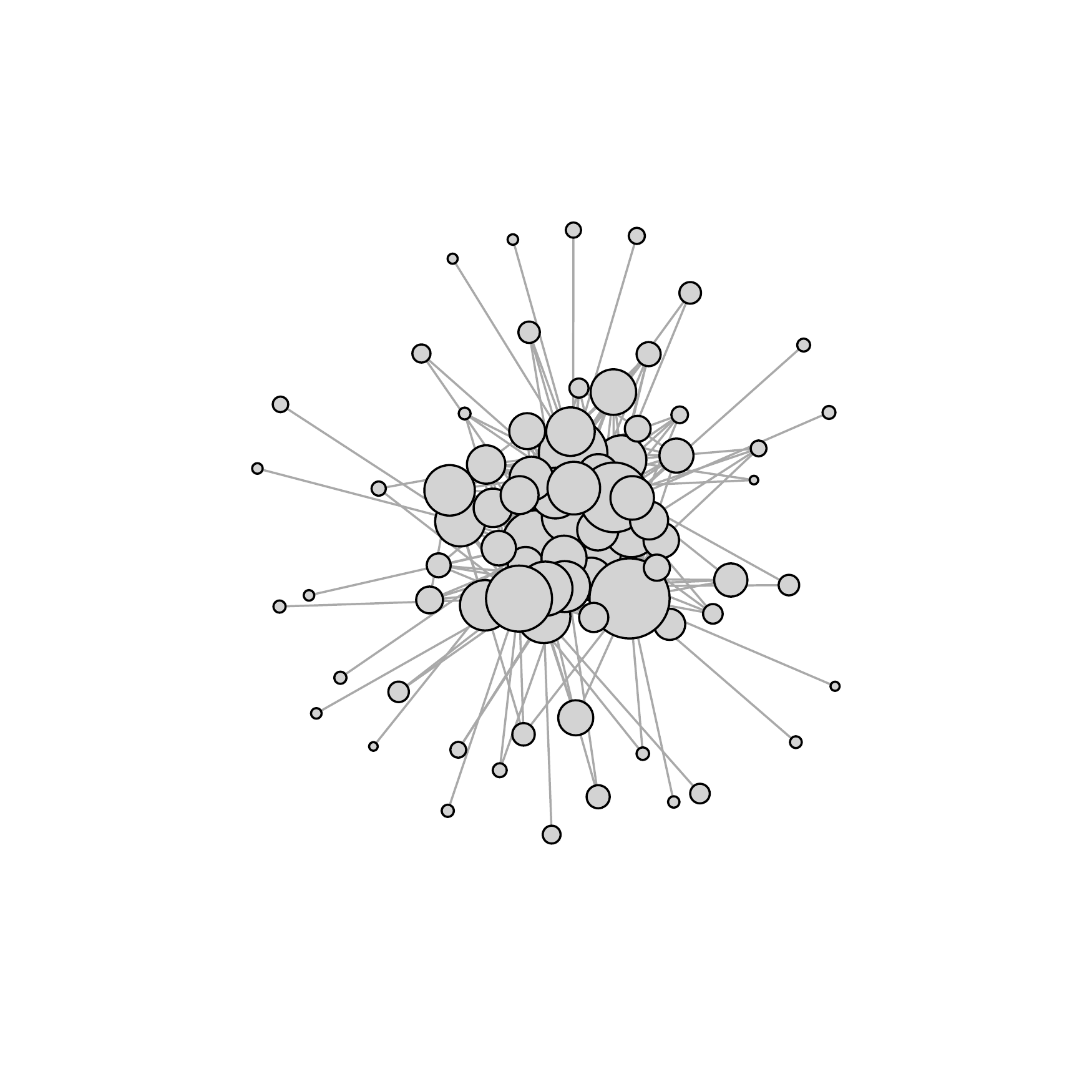}
\end{center}
\caption{\label{fig:communityGraphs} Community graphs of the \texttt{PGPgiantcompo} web of trust for (top to bottom) \textsf{PLP}, \textsf{PLM}, \textsf{PLMR} and \textsf{EPP(4, PLP, PLMR)}}
\end{figure}

However, we give an example to illustrate differences between our algorithms in a more qualitative way.  Coarsening the input graph according to the detected communities yields a 
\textit{community graph}, which we then visualize by drawing the size of nodes proportional to the size of the respective community.
Figure~\ref{fig:communityGraphs} shows community graphs for the \texttt{PGPgiantcompo} graph,
a social network and web of trust resulting from signatures on PGP keys.
From top to bottom, the solutions were produced by \textsf{PLP}, \textsf{PLM}, \textsf{PLMR} and \textsf{EPP(4, PLP, PLMR)}.
It is apparent that \textsf{PLP} has a much finer resolution and detects ca. 1000 small communities.
This is true for most of our data sets, but the inverse case also appears. 
On this network, higher modularity is associated with coarser resolution.
\textsf{PLM}, \textsf{PLMR} and \textsf{EPP(4, PLP, PLMR)} have a very similar resolution and divide the network into ca. 100 communities.
While \texttt{PGPgiantcompo} is admittedly a very small graph, this example shows how community detection can help to reduce
 the complexity of networks for visual representation.

\section{Conclusion and Future Work}

We have developed and implemented several parallel algorithms
for community detection, a common and challenging task in network analysis.
Successful techniques and parameter settings have been identified
in extensive experiments on synthetic and real-world networks. 
They include three standalone parallel algorithms, all of which are placed on the Pareto frontier with respect
to running time and modularity in an experimental comparison 
with other state-of-the-art implementations.
While the \textsf{PLP} label propagation algorithm is extremely fast, its solution might not
always be satisfactory for some applications. \textsf{PLM} is to the
best of our knowledge the first parallel variant of the established
Louvain algorithm which can handle massive inputs.
On our machine, it detects high-quality communities in a network with 3.3 billion edges in under $3$ minutes using $32$ threads. 
Achieving significant parallel speedups over the frequently used sequential algorithm, it can accelerate analysis workflows now and even further on future multicore systems.
Our modification \textsf{PLMR} of this method adds a refinement
phase which enhances modularity for a small increase in running time.

Our implementations are published as a component of \textit{NetworKit}~\cite{staudt2014networkit}, an open-source package 
of performant implementations for established and novel network analysis algorithms.
We invite researchers in algorithm engineering and network science to benefit from our software development efforts and consider contribution to the project.
\textit{NetworKit} is under active development by several researchers and may be extended by additional community detection methods in the future, \eg considering overlapping communities as well.

\bigskip

\begin{small}
\subsubsection*{Acknowledgements}
This work was partially supported by the project \textit{Parallel
Analysis of Dynamic Networks \textemdash{} Algorithm Engineering of
Efficient Combinatorial and Numerical Methods}, which
is funded by the Ministry of Science, Research and the Arts Baden-Württemberg.
We thank Pratistha Bhattarai for her contributions to the experimental study, 
Michael Hamann for optimizations to \algo{PLM}, and
numerous contributors to the \textit{NetworKit} project.
\end{small}

\medskip

\begin{scriptsize}
\copyright \ 2015 IEEE. Citation information: DOI10.1109/TPDS.2015.2390633, IEEE Transactions on Parallel and Distributed Systems. \url{http://ieeexplore.ieee.org/xpl/articleDetails.jsp?arnumber=7006796}
\end{scriptsize}

\newpage

\bibliographystyle{IEEEtranS}
\bibliography{Bibliography}

\bigskip

\begin{small}

\begin{wrapfigure}{l}{0.26\columnwidth}
\centering
 \includegraphics[width=2.6cm]{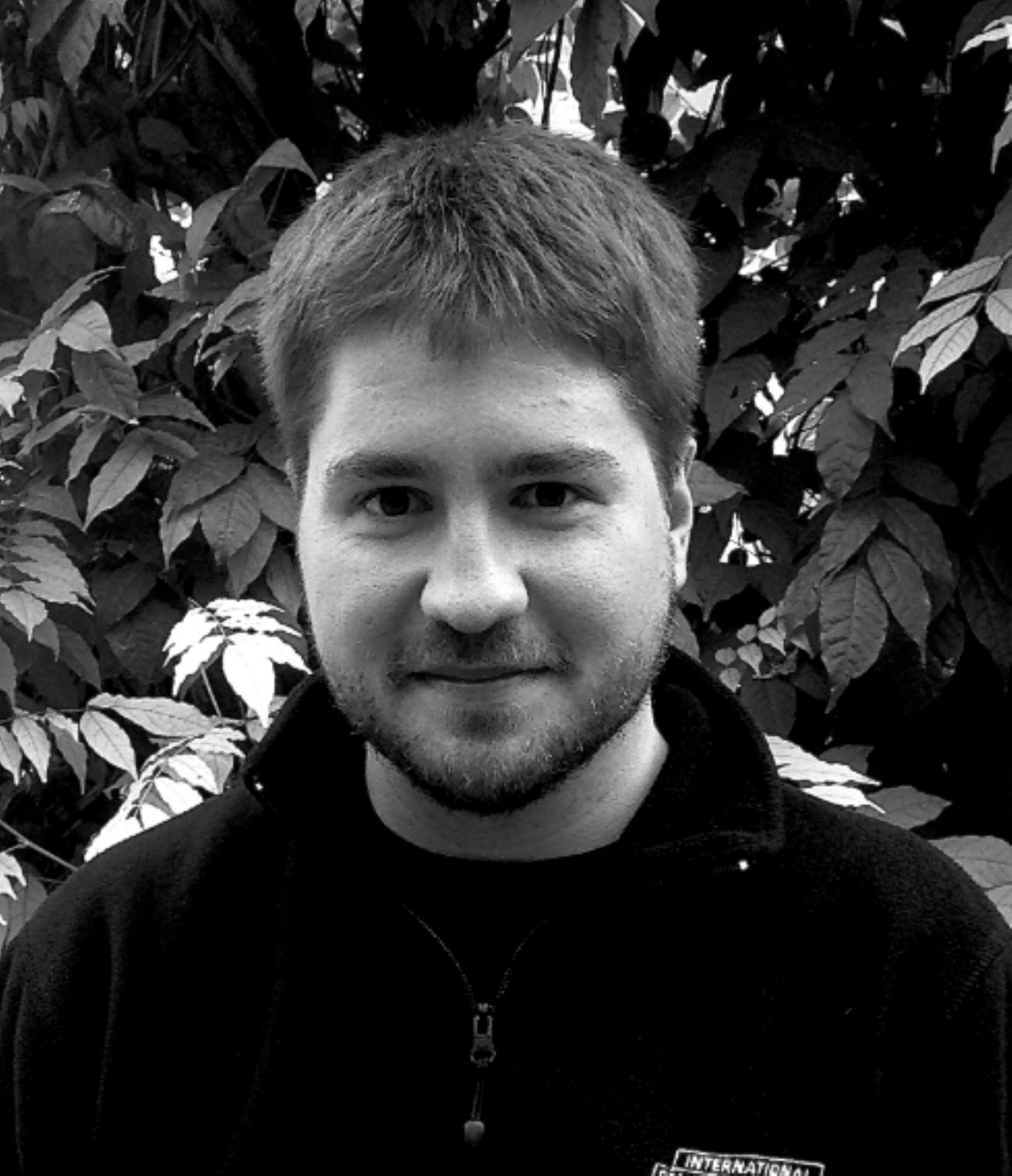}
\end{wrapfigure}

\noindent \textbf{Christian L. Staudt} received his Diplom degree in computer science from Karlsruhe Institute of Technology (KIT) in 2012. He is currently a researcher and PhD candidate in the Parallel Computing Group, Institute of Theoretical Informatics, KIT.  His research focuses on developing efficient algorithms and software for the analysis of large complex networks. Beyond that, he is interested in how network analysis methods can enable the study of complex systems in various domains. 

\medskip

\begin{wrapfigure}{l}{0.26\columnwidth}
\centering
 \includegraphics[scale=0.18]{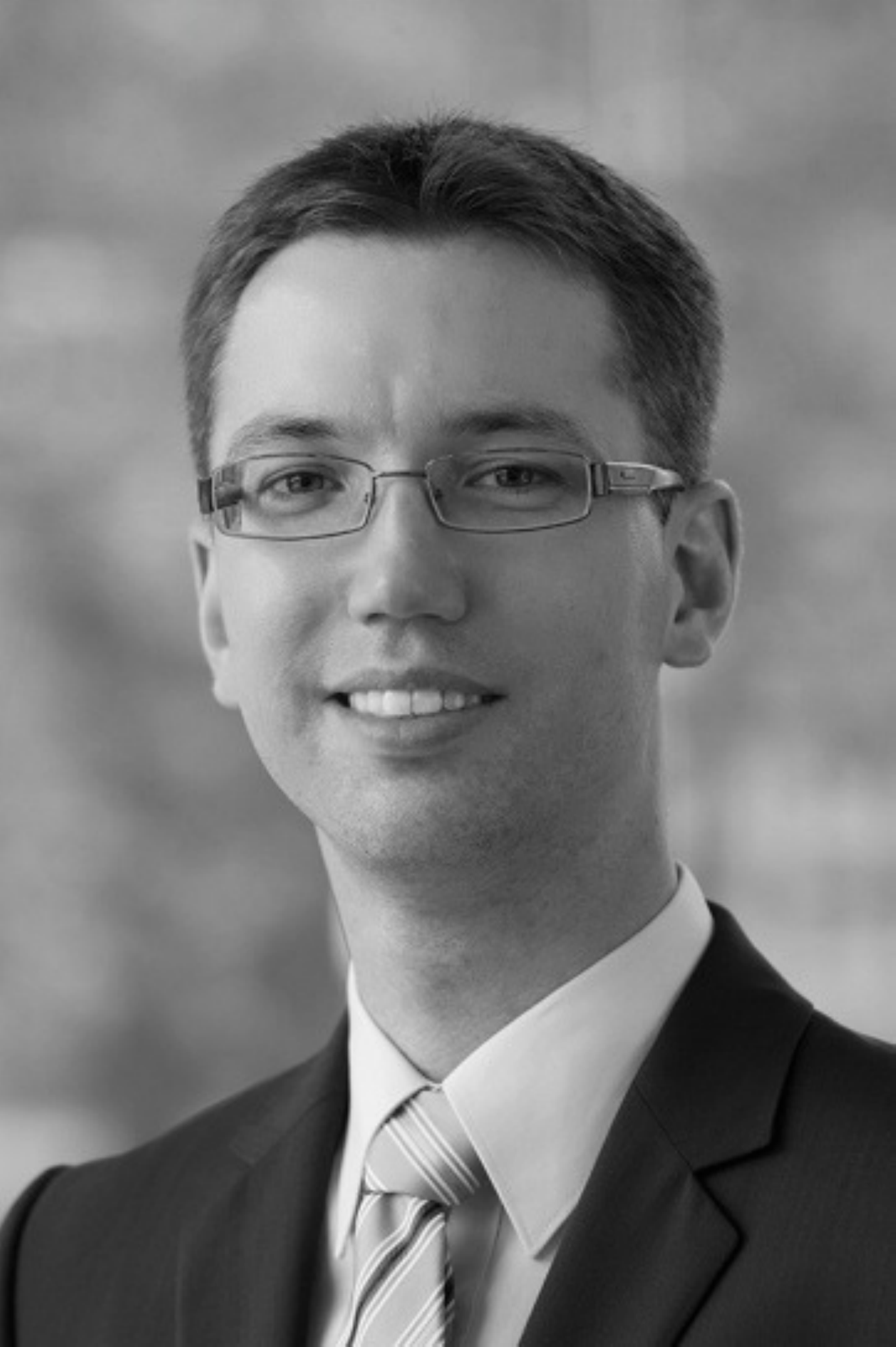}
\end{wrapfigure}

\noindent \textbf{Henning Meyerhenke}
is an Assistant Professor (Juniorprofessor) in the Institute of
Theoretical Informatics at Karlsruhe Institute of Technology (KIT), Germany, 
since October 2011. Before joining KIT, Henning was a Postdoctoral Researcher
in the College of Computing at Georgia Institute of Technology (USA) and at the University
of Paderborn (Germany) as well as a Research Scientist at NEC Laboratories Europe.
Henning received his Diplom degree in Computer Science from Friedrich-Schiller-University
Jena, Germany, in 2004 and his Ph.D. in Computer Science from the University of 
Paderborn, Germany, in 2008. Dr. Meyerhenke’s main research interests are efficient sequential
and parallel algorithms and tools for applications in network analysis, combinatorial 
scientific computing, and the life sciences.

\end{small}

\newpage
\newpage
\appendix

\section{Additional Experimental Results}

\begin{figure}[htb]
\begin{centering}
\includegraphics[width=0.901\columnwidth]{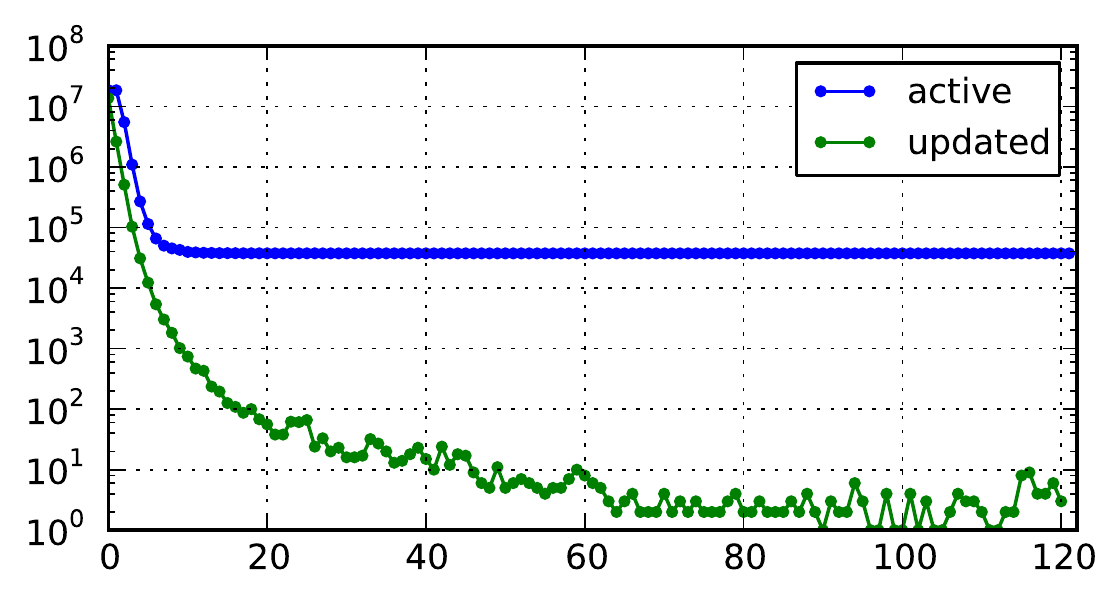}
\par\end{centering}
\caption{\label{fig:lp-update}Number of active and updated labels per iteration
of \textsf{PLP} for the web graph \textsf{uk-2002}.}
\end{figure}

\begin{figure}[htb]
\begin{centering}
\includegraphics[width=0.85\columnwidth]{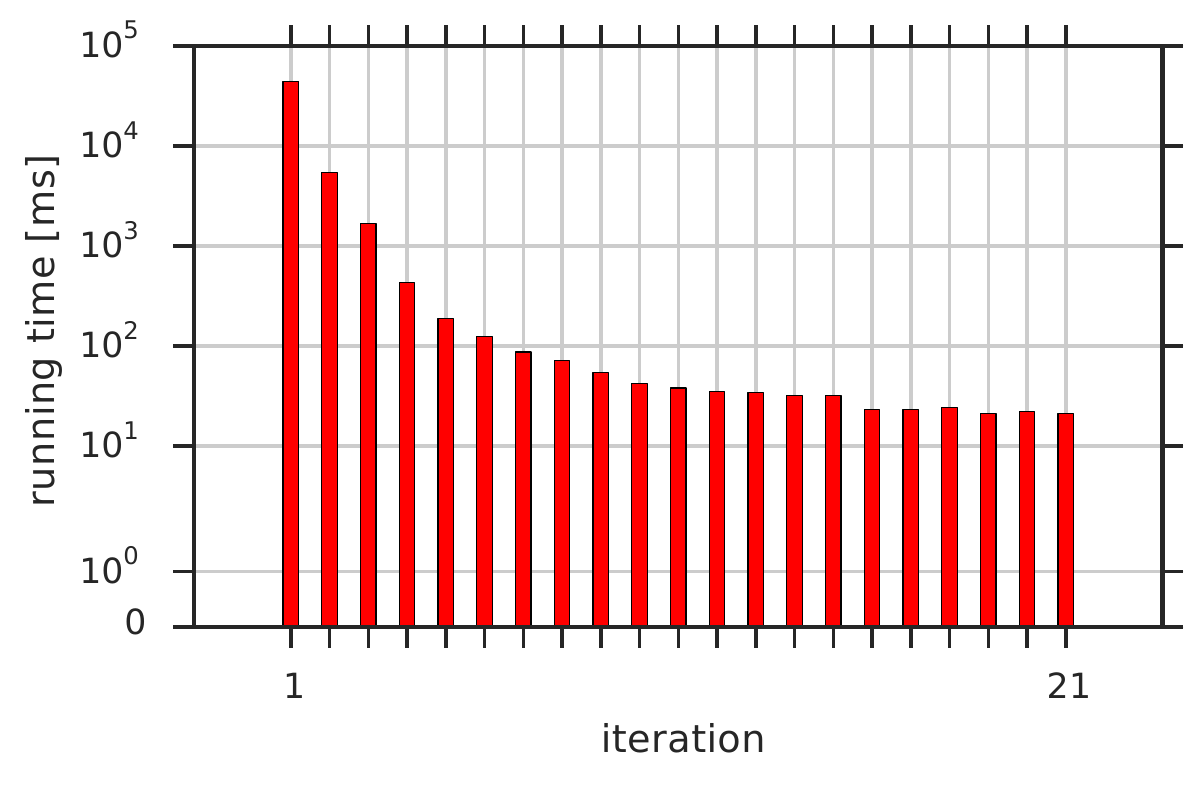}
\par\end{centering}
\caption{\textsf{PLP} running time in milliseconds per iteration for the \texttt{uk-2007-05} web graph, at 32 threads.} 
\label{fig:plp-iter}
\end{figure}

\begin{figure}[h]
\begin{centering}
\includegraphics[height=3.2cm]{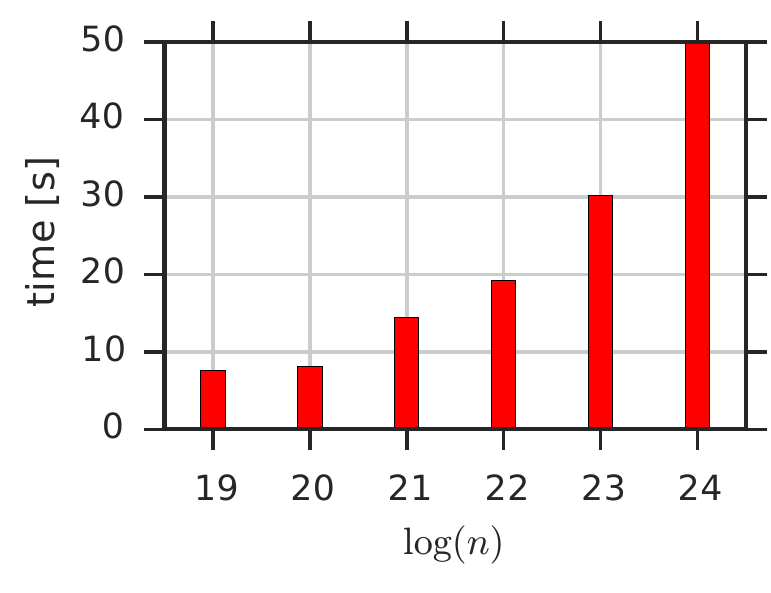}
\includegraphics[height=3.2cm]{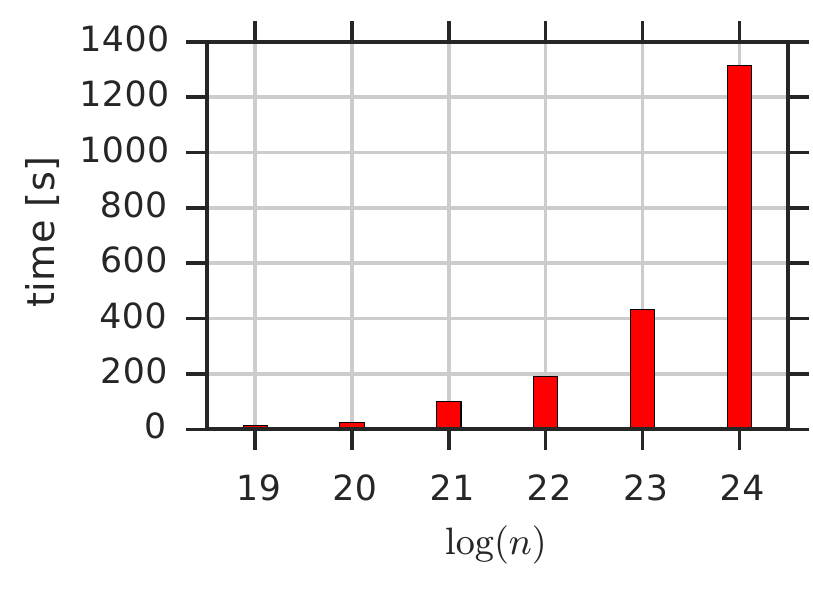}
\caption{\label{fig:weakscaling-rmat}
\textsf{PLP} (left) and \textsf{PLM} (right)
weak scaling on the series of R-MAT graphs.}
\end{centering}
\end{figure}

\onecolumn

  \begin{figure}[!p]
  \begin{center}
\includegraphics[width=.45\columnwidth]{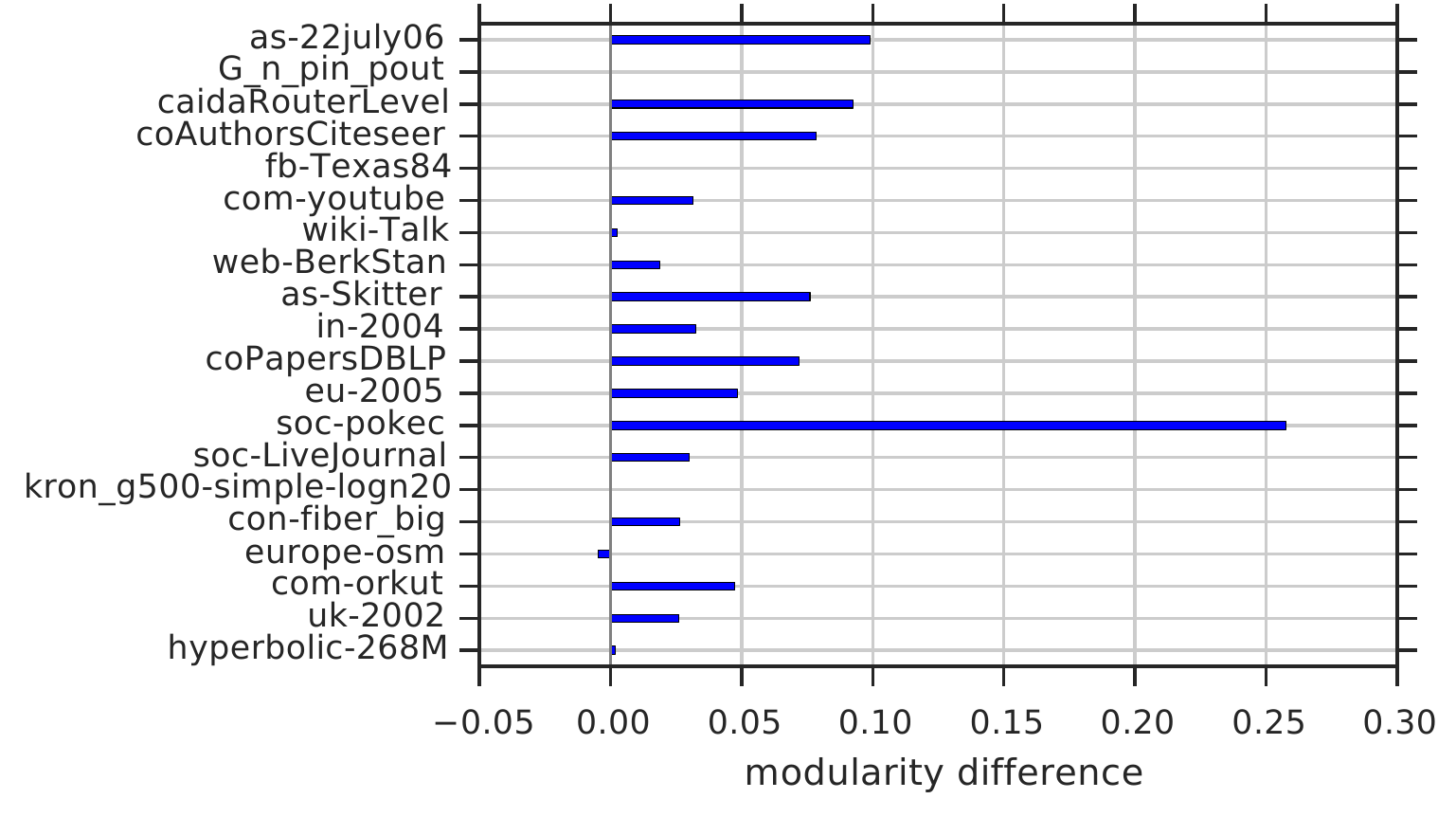}
\includegraphics[width=.45\columnwidth]{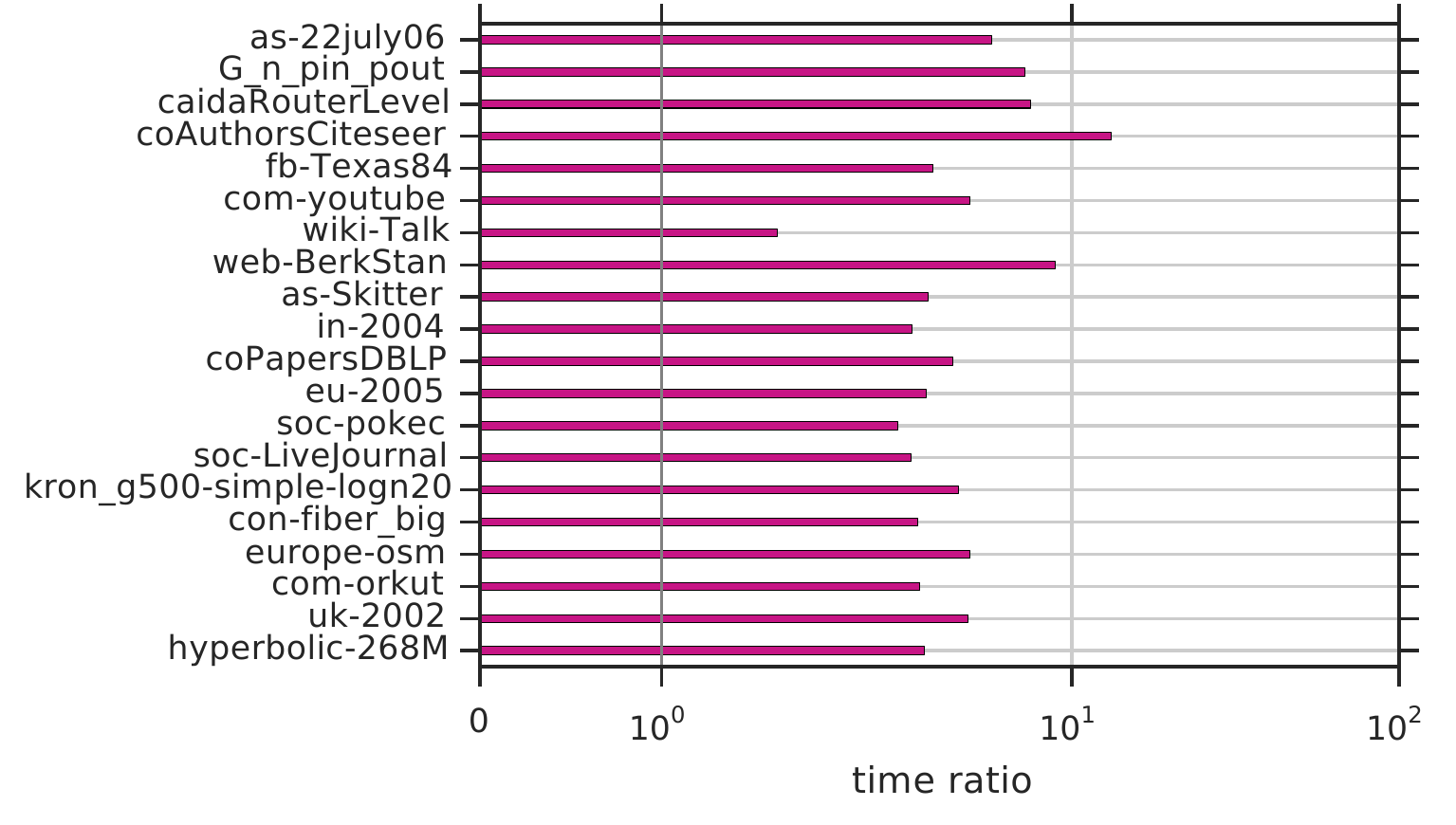}

\caption{\label{fig:EPP-vs-PLP} Difference in quality (left) and running time
time ratio (right) of \textsf{EPP(4,PLP,PLMR)} compared to a single \textsf{PLP}.} 
    \end{center}
  \end{figure}

  \begin{figure}[!p]
  \begin{center}
\subfloat[\textsf{CGGCi} \label{fig:cggci-rel}]{%
\includegraphics[width=.45\columnwidth]{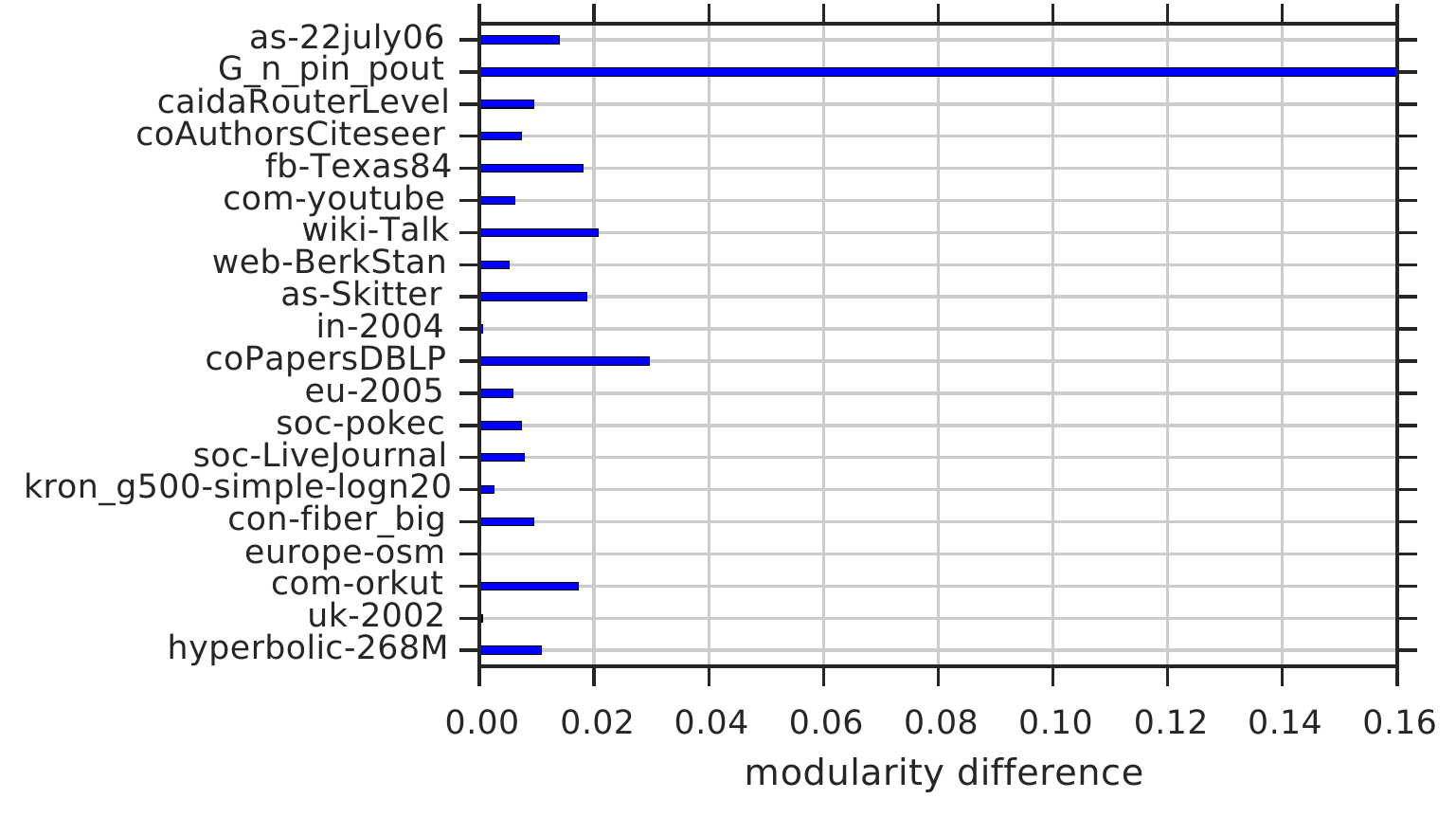}
\includegraphics[width=.45\columnwidth]{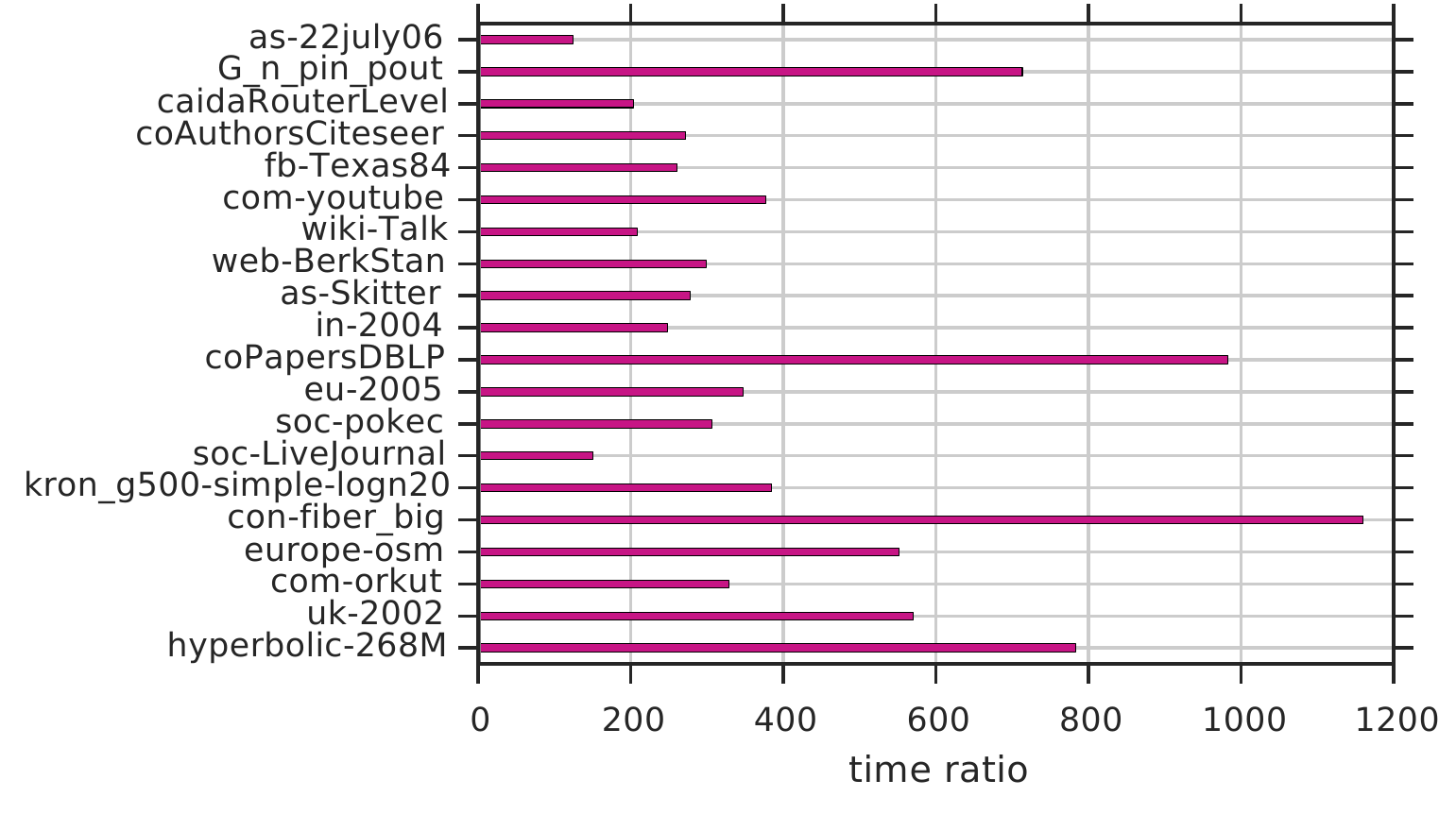}
    }
   \hfill
    \caption{Performance of the competitor algorithm CGGCi relative to baseline \textsf{PLM}.}
    \label{fig:performance-theirs2}
    \end{center}
  \end{figure}
  
    \begin{figure}[!p]
  \begin{center}
\subfloat[\textsf{PLM*} \label{fig:plm_}]{%
\includegraphics[width=.45\columnwidth]{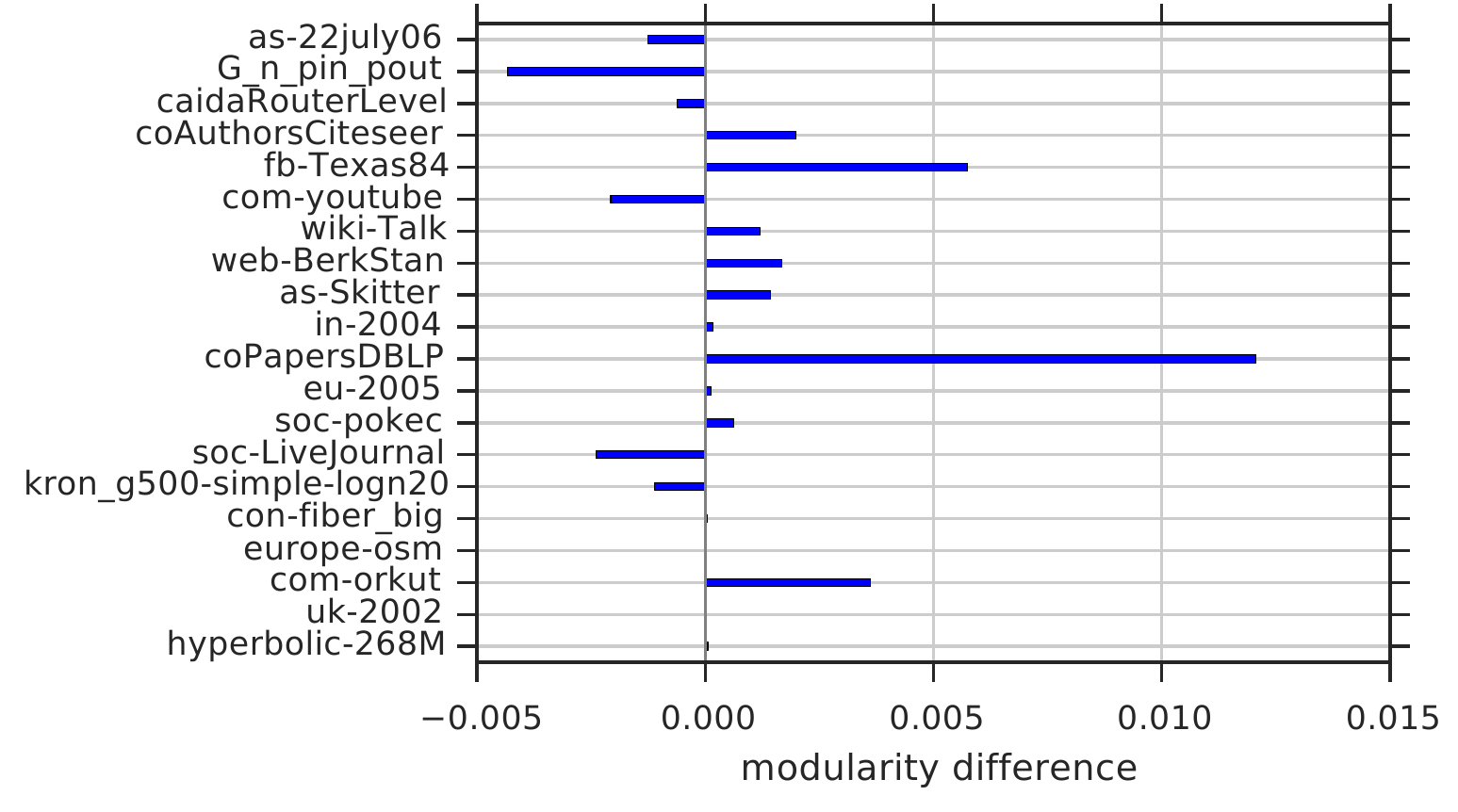}
\includegraphics[width=.45\columnwidth]{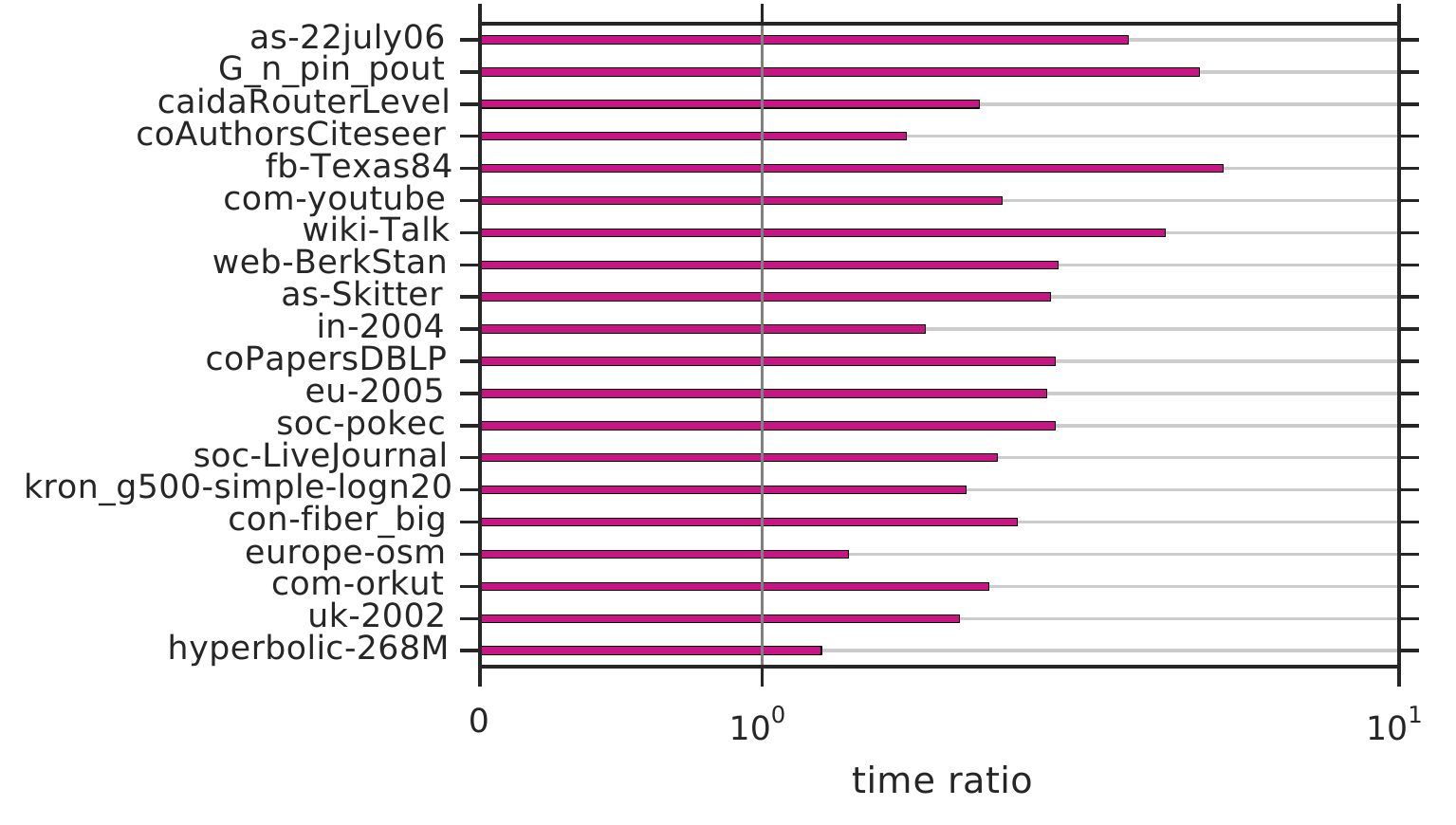}
    }
   \hfill
    \caption{Performance of our PLM* algorithm relative to baseline \textsf{PLM}.}
    \label{fig:performance-ours2}
    \end{center}
  \end{figure}

\end{document}